\documentclass{article}

\begin{document}

\def\giorno{26 June 2002}

\def\sn{**} 

\def\pa{\partial}
\def\a{\alpha}
\def\b{\beta}
\def\la{\lambda}
\def\s{\sigma}
\def\De{\Delta}
\def\Ga{\Gamma}
 
\def\I{{\cal I}}
\def\F{{\cal F}}
\def\L{{\cal L}}
\def\M{{\cal M}}
\def\Y{{\cal Y}}
\def\X{{\cal X}}
\def\V{{\cal V}} 
\def\W{{\cal W}}
\def\Y{{\cal Y}}
\def\Z{{\cal Z}}
\def\H{{\cal H}}
\def\h{{\cal H}}
\def\G{{\cal G}}

\def\sse{\subseteq} 
\def\ss{\subset}

\def\ker{{\rm Ker}} 
\def\ran{{\rm Ran}} 

\def\({\left(}
\def\){\right)}
\def\[{\left[} 
\def\]{\right]} 

\def\~#1{{\widetilde #1}}
\def\^#1{{\widehat #1}}
\def\=#1{{\widetilde #1}}
\def\frac#1#2{{#1 \over #2}}

\def\hot{{\rm h.o.t.}}

\def\eb{{\bf e}}
\def\vb{{\bf v}}
\def\xb{{\bf x}}

\def\C{{\bf C}}
\def\N{{\bf N}}
\def\R{{\bf R}}
\def\Q{{\bf Q}}

\title{Poincar\'e and Lie renormalized forms for regular singular points of vector fields in the plane}
 
\author{Giuseppe Gaeta\footnote{e-mail: {\it g.gaeta@tiscali.it}.} \\ 
{\it Dipartimento di Matematica, Universit\`a di Milano,} \\
{\it v. Saldini 50, I--20133 Milano (Italy)} }
 
\date{\giorno}

\maketitle

\noindent
{\bf Summary.} We discuss the local behaviour of vector fields in the plane $\R^2$ around a regular singular point, using recently introduced reduced normal forms, i.e. Poincar\'e and Lie renormalized forms \cite{LMP,IHP,GaL}. We give a complete classification, and provide explicit formulas, using Poincar\'e renormalized forms for non-degenerate cases, and Lie ones for certain degenerate cases. Both schemes are completely algorithmic, prove to be easy to implement, and only require to solve linear equations.

\section{Introduction}
\def\sn{1}

The theory and method of normal forms (NF)  \cite{Arn1,Arn2,CGs,Elp,Gle,IoA,Ver,Wal}, whose
origins go back to the work of Poincar\'e at the end of XIX century, constitute a fundamental tool to study the behaviour of dynamical systems locally near a known solution.
 
The normal form of a dynamical system ${\dot x} = f (x)$ (or equivalently of the vector field $X = f^i (x) \pa / \pa x^i$) with a fixed point in the origin depends on the properties of the matrix $A = (Df)(0)$. 
As well known, when the eigenvalues are nonresonant the normal form is just given by the linear part of the system, and thus unique, while for resonant eigenvalues the normal form is in general not unique and can depend on infinitely many arbitrary constants. 

Needless to say, this richness of the normal forms unfolding reflects the richness of possible different behaviours of nonlinear systems sharing the same linear part; however, it is well known that this is to some extent redundant: a given system with resonant linear part does not have a unique normal form, if not in exceptional cases (this is related to the freedom in choosing solutions to the homological equation at resonant orders $k$). 
Thus several authors have tried to devise ways to reduce this
redundancy of the normal form classification, and on the other side to take advantage of the freedom mentioned above; in this respect one should quote \cite{AFG,Bai,BaC,BaS,Bro,BrT,Brus,CDD,Kum,vdM,Tak,Ush} as well as \cite{San1,San2}. This problem was actually already mentioned by Dulac \cite{Dul}.
 
One of these attempts, proposed in \cite{LMP,IHP}, is based on considering higher order homological operators and the related homological equations (details on this approach will be given below); the resulting ``further simplified'' normal form has been called {\bf  Poincar\'e renormalized form } (PRF). In a later contribution \cite{GaL} it was noted that the PRF approach does not make use of the Lie algebraic structure of vector fields in normal form, and that in certain cases one can obtain a stronger simplification by considering this structure\footnote{Doing so is so natural that the structure was inadvertently used in some examples considered in \cite{LMP,IHP}; this is the content of the errata to these papers; see also section 10 below.}. The resulting ``further simplified'' normal form has been called {\bf Lie renormalized form } (LRF).

In this note I want to use PRFs and LRFs to analyze the behaviour of vector fields (dynamical systems) in the plane $\R^2$ locally near regular singular points (equilibria).

Together with general results, I will also give detailed computations up to some finite order (typically up to terms of order six in the $x,y$ variables) with explicit identification of the transformations needed to take a system in PRF, including closed-form expression of the numerical coefficients. This will show that the required computations are actually easy to implement in practice.
The analysis will be on the formal level only, which is standard in normal forms theory, and however useful for the analysis of the system in a way I will not discuss here (see e.g.
 \cite{Arn1,Arn2,CGs,Gle,IoA,Ver,Wal} for this matter). I will however show how one can explicitely determine the radius of convergence of the normalizing transformation at any finite order, see section 7.5.
 
Some confusion arised in the literature about PRFs, in particular concerning a system considered by Bruno and Petrovich and the application of the PRF scheme to this; the appendix is devoted to  clarifying this issue.

\medskip

We use frequently the abbreviations {\bf NF}(s) for normal form(s),  {\bf PRF}(s) for Poincar\'e renormalized form(s), and {\bf LRF}(s) for Lie renormalized form(s). Equations are consecutively numbered in each section, and we omit the section number when referring to equations of the same section. 

The financial support of {\it Fondazione CARIPLO per la ricerca scientifica} during part of this work is gratefully acknowledged.

\section{Poincar\'e normal and renormalized forms}
\def\sn{2}

Let us start by introducing some general notation. 
We will consider vector fields in $\R^2$; we will use coordinates $(x,y)$ in $\R^2$, corresponding to the basis $(\eb_1 , \eb_2)$, and denote a generic vector as $\xi = (\xi^1 , \xi^2)$. We will also write $\pa_i \equiv \pa / \pa \xi^i$.
 
We will denote by $\F$ the set of polynomial vector functions, i.e. of
polynomial functions $f : \R^2 \to \R^2$, having a zero in the origin; we denote by $\F_k \ss \F$ ($k \ge 0$) the set of polynomial vector functions homogeneous of degree $k+1$ in the $\xi$.
 
We denote by $\W$ the Lie algebra of polynomial vector fields in $\R^2$ equipped with the commutator operation.
 
If we focus on the coordinate expression of vector fields, the role of the commutator is taken by the Lie-Poisson bracket $\{ . , . \}$ defined as 
$$
\{ f,g \} \ := \ (f^j \pa_j ) g - (g^j \pa_j ) f \ . \eqno(\sn.1) $$
Indeed,
writing $X_f = f^i \pa_i$, $X_g = g^i \pa_i$, we have
$$ [ X_f , X_g ] \ = \
X_{\{f,g\} } \ . \eqno(\sn.2) $$
 
The set $\F$ equipped with the bracket $\{.,.\}$ is a Lie algebra. 
Notice that $\{ ., . \} : \F_k \times \F_m \to \F_{k+m}$.

We will also, with an abuse of notation, denote by $\W_k$ the set of vector fields whose components are homogeneous of degree $k+1$ in the $\xi$, and by $W_k$ the homogeneous part of order $k+1$ of the vector field $W$. Obviously
these are not intrinsically defined notions, but depend on the coordinates we use; thus if we consider a vector field $W$, when we change coordinates the $W_k$ will also change (but near identity changes of coordinates $\xi^i \to \=\xi^i = \xi^i + \psi^i (\xi)$, where $\psi \in \F_m$, will preserve the $W_k$ with $k < m$).

To the linear part $A \xi$ of a vector function $f \in \F$ we associate the homological operator $\L_0 = \{ A \xi , . \}$.
Notice that $\L_0 : \F_k \to \F_k$. 

We can also define the homological operator in $\W$ rather than in $\F$, as follows. If the linear part of $f$ is given by $A\xi$, we will denote the linear part (in the $\xi$ coordinates) of
$X_f$ as $X_A$. To the linear part $X_A$ of a vector field $X_f$ we associate the homological operator $\L_0 = [X_A , . ]$; note that $\L_0 : \W_k \to \W_k$.

We will equip $\F_k$ (and thus all of $V = \F_0 \oplus \F_1
\oplus....$) with the Bargmann scalar product \cite{Elp,IoA}; this is defined as follows:
 $$ ( x^{\mu_1} y^{\mu_2} \eb_\a \, , \, x^{\nu_1} y^{\nu_2}
\eb_\b ) \ := \
 \delta_{\a,\b} \, \langle \mu , \nu \rangle \ := \delta_{\mu
, \nu} \ {
 \pa^{\mu_1 + \mu_2}\, x^{\nu_1} y^{\nu_2} \over \pa x^{\mu_1} \,
\pa y^{\mu_2} }
 \eqno(\sn.3) $$
With this choice\footnote{Using the standard scalar product \cite{Arn1} would differ, here and below, only in some coefficients.} of scalar product in $V$, the adjoint of $\L_0$ is given by
$\L_0^+ = \{ A^+ \xi , . \} $, where $A^+$ is the adjoint of $A$: $A^+_{ij} = A^*_{ji}$.
   
The operators $\L_0$ and $\L_0^+$ play a crucial role in discussing the properties of $f$ under Poincar\'e transformations, i.e. under near-identity changes of coordinates in $\R^2$,
 given by $ {\=\xi}^i = \xi^i + h^i (\xi )$, with $ h \equiv h_k
\in \F_k$. 
  
\subsection{Standard Poincar\'e normal forms}

It is well known that by a careful use of Poincar\'e transformations, i.e. performing them for $k=1,2,...$ successively and
choosing the $h_k$'s as solution to the homological equations (see below), one can eliminate all terms in the range of $\L_0$.

That is, one can pass to coordinates $\eta$ which
reduce the coordinate expression of $f$ to a form (the {\it Poincar\'e-Dulac normal form}, or simply normal form) ${\^f}$, where $\^f (\eta) = A \eta + \^F
(\eta )$ and $\^F \in \ker (\L_0^+)$. 
 
It is also well known that $\ker (\L_0^+)$ belongs to (and coincides with for $A$ semisimple) the set of resonant vectors, which are defined as follows. Consider a basis in $\R^2$ such that $A$ is in Jordan normal form, and let $\la_1 , \la_2$ be its eigenvalues (possibly equal).

Then a resonant monomial vector $\xi^\mu \eb_\b$ of order $|\mu|$
is a vector $\vb$ with components (we write the
vector indices as lower ones for ease of notation) $\vb_\a = \xi_{(1)}^{\mu_1}
\xi_{(2)}^{\mu_2} \delta_{\a,\b} = x^{\mu_1} y^{\mu_2} \delta_{\a,\b}$, 
where $|\mu| = \mu_1 + \mu_2 > 1$ and the
$\mu_i$ are non-negative integers satisfying the {\it resonance relation}
$$ \mu_1 \la_1 \, + \, \mu_2 \la_2 \ = \ \la_\b \ . \eqno(\sn.4) $$
The linear span of resonant monomial vectors is the space of resonant vectors, i.e. $\ker (\L_0^+) \backslash [ \ker (\L_0^+) \cap \F_0]$.

The {\it homological equation} for $h_k$
is given as follows: let $\~f$ be the expression of $f$ obtained after
operating the previous Poincar\'e transformations, and let $\pi_k$ be the projection operator $\pi_k : \F_k \to \ran (\L_0) \cap \F_k$; then 
$$ \L_0 (h_k ) \ = \ \pi_k \, {\~f}_k \eqno(\sn.5) $$
is the required homological equation for $h_k$; notice that the solution to this is uniquely defined up to elements of $\ker (\L_0 )$.

We refer e.g. to \cite{Arn1,Arn2,CGs,Elp,Gle,IoA,Ver,Wal,Wal3} for
further detail on standard normal forms and the normalizing transformation.

\subsection{Poincar\'e renormalized forms}

In order to discuss PRFs, it is convenient to use Lie-Poincar\'e --  rather than Poincar\'e -- transformations. Let us first of all  briefly discuss these, referring to e.g. \cite{BGG,Dep,MiL,Wal3} or \cite{CGs,IHP} for further detail.
 
The function $h : \R^2 \to \R^2$, or
equivalently  the vector field $H = h^{(1)} (x,y) \pa_x + h^{(2)} (x,y) \pa_y$,
generates a Lie-Poincar\'e transformation given by the time-one flow of $H$.
Thus under the Lie-Poincar\'e transformation generated by the vector field
$H$, the vector field $W$ is transformed into
 $$ {\widetilde W}  =  e^H W
e^{-H}  ; \eqno(\sn.6) $$
this can be computed up to any desired order by
means of the
classical Baker-Campbell-Haussdorff formula as 
$$ {\widetilde
W} \ = \ \sum_{s=0}^\infty {1 \over s!}  [[ H , W ]]^s  ,
 \eqno(\sn.7) $$
where we have defined the iterated commutators as 
 $$ [[H,W]]^0 := \ W \ \ ; \ \ 
[[H,W]]^s := \ \[ \, H \, , \, [[H,W]]^{s-1} \, \] \ (s\ge 1) \ .
 \eqno(\sn.8) $$
  
If $h = h_k
\in \F_k$, from the above we have, denoting by $[a]$ the
integer part of
 $a$ and with $\h (f) := \{ h , f \}$, that 
$$ {\widetilde f}_m  \ = \   
\sum_{s=0}^{[m/k]}  {1 \over s!} \, \h^s (f_{m-sk} )
\ . \eqno(\sn.9) $$

We can now define Poincar\'e renormalized forms (PRFs) and
describe their construction.

We define the higher homological operators $\L_k$ as $\L_k := \{ f_k , . \}$; note that these make good sense only after $f_k$ has been stabilized in the procedure, as discussed in \cite{LMP,IHP}.
 
We define the spaces $H^{(p)} \sse \F$ ($p\ge 0$) by
$H^{(0)} = \F$, and $H^{(p+1)} = H^{(p)} \cap \ker (\L_p)$ for $p\ge 0$. 
This implies that $H^{(p+1)} \sse H^{(p)}$, and 
$$ H^{(p)} \ = \ \ker (\L_0 ) \cap ... \cap \ker (\L_{p-1} ) \ \equiv \ {\bigcap }_{s=0}^{p-1} \, \ker (\L_s ) \ . \eqno(\sn.10) $$
 
The restriction of $\L_p$ to $H^{(p)}$ will be denoted as $\M_p$. 
With this definition, we have $H^{(p+1)} = \ker (\M_p)$.
 
We also define the spaces $F^{(p)} \sse \F$ ($p\ge 0$) as $F^{(0)} = \F$ and $F^{(p)} = F^{(p-1)} \cap \ker (\M_p^+ )$ for $p \ge 1$. This implies that $F^{(p+1)} \sse F^{(p)}$, and
$$ F^{(p)} \ = \
{\bigcap }_{s=0}^p \, \left[ \ran (\M_s ) \right]^\perp \ = \ {\bigcap
}_{s=0}^p \, \ker ( \M_s^+ ) \ . \eqno(\sn.11) $$
(the orthogonal complement must be understood in $\F$ equipped with the scalar product). We also have $F^{(p+1)} =
F^{(p)} \backslash [\ran (\M_p) \cap F^{(p)} ]$.
 
We can also define the projection operators $\pi_k : \F \to \ker (\L_k )$, and $\Pi_s = \pi_{s-1} \circ ... \circ \pi_0$ for $s>0$, $\Pi_0$ the identity operator. Similarly, we define the projection operators  $P_s : \F \to \ran (\M_s )$. Notice that with these $H^{(p)} = \mu_p \F$, $\M_p = \L_p \circ \mu_p$.
 
The function $f \in \F$ (the associated vector field $X_f \in \W$) is said to be in PRF if $f_k \in F^{(k)}$ (and then obviously $f^k \in F^{(k)}_k :=  F^{(k)} \cap \F_k$). 

It can be shown that any function $f \in \F$ (any vector field  $W \in \W$) can be taken to PRF by a sequence of suitably chosen Lie-Poincar\'e transformations. 
  
Let us now briefly describe two possible schemes for constructing the sequence of ``suitably chosen'' transformations; these were discussed in \cite{LMP,IHP}.
\medskip

{\bf (A).} In the first case, denote by $f_k^{(0)}$ the term obtained after completing the procedure up to  order $k-1$. Then operate a series of transformations with generators\footnote{The lower index keeps track of the subspace $\F_k$ to which $h$ belongs; the upper index keeps track of the number of transformations already operated.} $h_k^{(0)} , h_{k-1}^{(1)}, ... , h_1^{(k-1)}$, with $h_p^{(s)} \in H^{(s)} \cap \F_p$. These should be chosen as solutions to the higher order homological equations
$$ P_s f_k^{(s)} \ = \ \M_s \left( h_{k-s}^{(s)} \right) \ ; \eqno(\sn.12) $$
in other words,
$$ h_{k-s}^{(s)} \ = \ \Pi_s \circ \M_s^+ \circ P_s ( f_k^{(s)} ) \ .
\eqno(\sn.13) $$  

{\bf (B).} Rather than putting $f_k^{(s)}$ in $F_k^{(s)}$ for $s=1,2,...,k$, and doing this for all $k=1,2,...$, we can invert the order of iterations, i.e. put $f_k^{(s)}$ in $F_k^{(s)}$ for all $k \ge s$, and do that for all $s=1,2,...$; in this case for $s=1$ we obtain the standard NF. 
\medskip

Notice that the equations to solve, and the spaces to which the functions belong, are the same in the two cases; however, the form to which $f_k$ has been taken by previous parts of the procedure when we deal with $f_k^{(s)}$ can be different.

Due to non-unicity of PRF, these two procedure can indeed in principles give different PRFs, i.e. the arbitrary coefficient which appear in the general form of PRFs for a given system can take different values depending on the procedure we have followed. 

The reader is referred to \cite{CGs,IHP} for further detail concerning Poincar\'e renormalized forms and related matters, including the Hamiltonian version of the theory and the role of (linear) symmetries.

We stress that the schemes mentioned here are ``generic'', i.e. do not take into account the Lie algebraic structure of the set $\G$ of vector fields in normal form (with respect to a given linear part). This point will be considered in section 3 below, where a ``$\G$-adapted'' procedure is discussed; this will also make transparent the relation between this approach and Broer's one \cite{Bro}.\footnote{I would like to stress that the idea of using $\L_k$ with the same role as $\L_0$ was already contained in \cite{Tak}; at the time of writing \cite{LMP,IHP} I had not realized this, and did not give proper credit.}

\section{Reduction of normal forms and Lie algebras}
\def\sn{3}

The set of vector fields in (standard) normal form with respect to a given linear part has a natural Lie algebraic structure.
It turns out that making use of this, one can obtain a better reduction of the normal form than using only the graded structure.

\subsection{The LRF approach}

Consider the set of vector fields in $\R^n$ which are in normal form with respect to the given linear part $A$, i.e. the vector fields $Y$ such that $[X_A , Y ] = 0$. It is obvious that these form a Lie algebra (the Lie operation being the standard commutator of vector fields); we denote this algebra by $\G_A$.
 
Let us recall a general characterization of vector fields in normal form relevant in this context \cite{CGs,Elp,IoA,Wal}. Consider the linear vector field $X_A$; we say that the differentiable function $\phi : \R^n \to \R$ is an invariant for $X_A$ if $X_A (\phi ) = 0$. 

We will denote by $\I^* (A)$ the set of invariants for $X_A$ which are meromorphic (i.e. a quotient of algebraic functions) in the $x$ coordinates; by $\I (A) \ss \I^* (A)$ the set of algebraic invariants for $X_A$; and by $\I_k (A) \ss \I (A)$ the set of algebraic invariants for $X_A$ which are homogeneous of degree $k+1$ in the $x$ variables.

Let $G = C(A)$ be the centralizer of $A$ in the group $GL(n)$; let its Lie algebra be spanned by matrices $\{ K_1 , ... , K_d \}$ (we can always assume $K_1 = I$, and that $K_\a = A$ for some $\a$, provided $A \not= 0$; notice that $d \le n$). The vector fields corresponding to these are given in the $x$ coordinates by $X^{(\a)} = (K_\a x)^i \pa_i$. 

The most general vector field $W$ in $\G$ can be written as
$$ W \ = \ \sum_{\a=1}^d \ \mu_\a (x) \ X^{(\a)}  \eqno(8)$$
where $\mu_\a (x) \in \I^* (A)$; this implies that $\G$ is contained in a finitely generated module over $\I^* (A)$. 

As $W$ must be algebraic in the $x$, and $X^{(\a)}$ are linear in $x$, functions $\mu_\a (x) \in \I^* (A)$ with poles of degree $d \ge 2$ in $x = 0$ cannot appear in (10). In other words, only algebraic functions and functions with simple poles in the origin can appear in the actual normal form unfolding; hence $\G$ is not, in general, the full $G$-generated module over $\I^* (A)$. 

In several cases, however, it happens that $\G$ has a more convenient structure, i.e. the $\mu_\a$ in (10) can actually be taken to be in $\I (A)$, and not just in $\I^* (A)$. In this case we say that all the vector fields in $\G$ are {\it quasi-linear}, or that we have a {\it quasi-linear normal form}. 
In particular, this is the case when $A$ admits only one basic invariant (as in the cases we are interested in).

If the normal form is quasilinear, we have $\G \cap \V_{k+1} = \I_k (A) \otimes G$, and the analysis of the structure of $\G$ results to be particularly simple.

Call $\X^*_\a$ the algebra spanned by vectors which are written as $X = \mu (x) X^{(\a)}$ with $\mu \in \I^* (A)$, and $\X_\a$ the algebra spanned by vectors with $\mu \in \I (A)$ (this is the module over $\I (A)$ generated by $X^{(\a)}$). In general we have $ \X_1 \oplus ... \oplus \X_d \subseteq \G \subset \ \X^*_1 \oplus ... \oplus \X^*_d$, and in the quasi-linear case we actually have $\G = \X_1  \oplus ... \oplus \X_d$.

It can happen that we are able to determine a sequence of subalgebras $\F_p \subseteq \G$, each of them being the union of $\X_\a$ subalgebras, such that $\F_0 = \G$ and 
$$ \[ \, \G \, , \, \F_p \, \] \ = \  \F_{p+1} \ ; \eqno(10) $$
if this terminates in zero we say that $\G$ has a {\it quasi-nilpotent} structure. Notice that the factor algebras $\Gamma_p := \F_p / \F_{p+1}$ are in general {\it not } abelian.

It results \cite{GaL} that $\G$ can have a quasi-nilpotent structure only if $G$ is nilpotent. The chain of subalgebras $\F_p \subset \G$ can then be read off the descending central series $G_p$ of $G$. 
The subalgebras $\Ga_p$ introduced above are therefore moduli over $\I (A)$ generated by the abelian subalgebras $\gamma_p = G_p / G_{p+1}$ of $G$.

If $\G$ is quasi-nilpotent, we can first work with generators in $\Ga_1$ and simplify terms in $\Ga_1$, then consider generators in $\Ga_2$ and simplify the corresponding terms being guaranteed that $\Ga_1$ terms are not changed, and so on. 

Notice that in this case we are -- roughly speaking --  just using the nilpotent structure of (the finite dimensional group) $G$, rather than the one of (the infinite dimensional algebra) $\G$.

{\bf Remark 6.} It should be stressed that the reduced normal form obtain in this way is {\bf not}  necessarily a PRF in the sense of the definition discussed in section 2. We will use therefore the name {\bf Lie renormalized form} (LRF) to emphasize that it is obtained using the Lie algebraic properties of the set of vectors in normal form and at the same time the main idea behind the PRF procedure. $\odot$

Needless to say, this approach is particularly convenient when the $\Ga_p$ are generated by a single element of $G$.

The situation depicted above is met in applications: in particular, it applies to any nontrivial two-dimensional case. More generally, it always applies when there is only one basic invariant.

A concrete application of this ``$\G$-adapted procedure'' will be given below when considering certain subcases, see subsections 6.3 and 7.4; in this case, indeed, the generic PRF procedure given in \cite{LMP,IHP} would produce an infinite PRF (as shown in subsections 6.1 and 6.2), while the $\G$-adapted one will produce a finite LRF (as shown in subsection 6.4). In this case, it will turn out that the LRF is not a PRF.

{\bf Remark 7.} It should also be emphasized that this procedure can be seen as an implementation of Broer's idea on reduction of normal forms as filtration of Lie algebras; see also the work of Baider and coworkers. However, it is {\it not } exactly following their general procedure, based on descending central series (see the discussion in appendix B of \cite{GaL}) but rather it is a blend of the Broer-Baider procedure and of the general PRF procedure. It should also be mentioned that the Broer approach is extremely general, and produces unique normal forms; see \cite{Bai,BaC,Bth,Bro} for discussion and applications, and \cite{San1,San2} for a clarification of the issue. However, implementing it in practice can be of considerable difficulty, and as far as I know explicit computations based on this have been performed only for very simple systems. 
On the other hand, the LRF approach is not as general as Broer's approach (it requires the algebra $\G$ to have a favourable structure) and will in general not produce unique reduced normal forms; but when applicable it only requires linear algebra computations and is easily implementable on a computer.
$\odot$

\section{Singular points of vector fields in the plane: the basic
classification of linear parts.}
\def\sn{4}

Let $A = (Df)(x_0)$ be the linear part of $X = f^i \pa_i$ at the equilibrium point $x_0$; we can and will always shift coordinates in $R^2$ so that $x_0$ is in the origin.

After reduction to Jordan normal form, and up to 
permutation of
coordinates, the following cases are possible for $A \not= 0$ (all the
constants $\mu , \mu_i$ below are understood to be real and nonzero, and $\s = \mu_1 + i \mu_2$):

$$
\begin{array}{llcll}
A = \pmatrix{\s & 0 \cr 0 & \s^* \cr} & (S1) & ; &
A = \pmatrix{i \mu & 0 \cr 0 & - i \mu \cr} & (S2) \\ 
A = \pmatrix{0 & 0 \cr 0 & \mu \cr} & (S3) & ; & 
A = \pmatrix{\mu_1 & 0 \cr 0 & \mu_2 \cr} & (S4) \\ 
A = \pmatrix{\mu & 1 \cr 0 & \mu \cr} & (N1) & ; & 
A = \pmatrix{0 & 1 \cr 0 & 0 \cr} & (N2) \end{array} $$
In cases {\bf S1}-{\bf S4} the matrix $A$ is semisimple, in case {\bf N1} it has a semisimple part and a nilpotent one, in case {\bf N2} it is nilpotent, with zero semisimple part. Thus, case {\bf N2} (and the case $A=0$) corresponds to a non-regular singular point \cite{Arn2}; it is known that in this case normal forms theory is not very effective \cite{Arn2,Wal}, and we will not deal with this; case {\bf N2} is studied from the point of view of PRFs in \cite{IHP}, but the results that can be obtained are very poor (this singularity is better studied with different methods, see \cite{Tak} and \cite{BaS,KOW}).

\bigskip
 
We will now briefly recall the results obtained by standard NF theory
for each of the cases listed above. In several of them the PRFs are 
either trivial (i.e. coincide with the standard NF) or have been studied in [1,2].
 
In the generic case {\bf (S1)} no resonance can be present (recall we assumed $\mu_i \not=0$), so the NF is linear; moreover the eigenvalues belong to a Poincar\'e domain, so the normalizing transformation is convergent in some sufficiently small neighbourhood of the origin \cite{Arn1,Arn2}.

The case {\bf (S2)}, which is generic for hamiltonian system, has infinitely many resonances. The NF is written as $W = [ 1 + \a (|x|^2 ) ] Ax + \b(|x|^2) Ex$, where $|x|^2 = x_1^2 + x_2^2$, $\a$ and $\b$ are arbitrary polynomial functions with zero constant part, and $E$ is the identity matrix; 
in the hamiltonian case we obviously have $\b \equiv 0 $. The further reduction of this NF has been considered by Siegel and Moser \cite{SiM} in the hamiltonian case (see \cite{FoM} for higher dimensions), while the generic case $\b \not\equiv 0$ has been studied via PRFs in \cite{LMP,IHP} (see section 10 below). Note that for $\b \equiv 0 $ the NF satisfies ``condition A'' and thus, provided the linear part satisfies the arithmetic ``condition $\omega$'' \cite{Bru,CGs}, the normalizing transformation is convergent on the basis of Markhashov-Bruno-Walcher-Cicogna theory  \cite{BrW,Cic1,Cic2,Cic3,CGs,Mar,Wal4}; no convergence result is available in the generic case.

In case {\bf {\bf S3}} the eigenvalues cannot belong to a Poincar\'e domain, and it is easy to see that the NF will depend on two infinite sequences of real constants, 
$$ \begin{array}{rl} {\dot x} =& \sum_{k=1}^\infty a_k x^{k+1} \\ 
 {\dot y} =& \sum_{k=1}^\infty b_k x^k y \ . \end{array} \eqno(\sn.1) $$
The PRF in this case will be studied in detail in sections 6 and 7 (see also the appendix).

In the case {\bf {\bf S4}} we should distinguish several subcases according to two criteria: first, if $\mu_1 / \mu_2$ is rational or irrational, and second according to the sign of $\mu_1 \mu_2$. 

For $\mu_1 \mu_2 > 0$ the eigenvalues belong to a Poincar\'e domain, and the transformation to NF is guaranteed to be convergent on the basis of the Poincar\'e criterion; if $\mu_1 / \mu_2$ is irrational, the NF is linear, otherways it can include resonant nonlinear terms (see subsection 8.1 for limitations on these). 

For $\mu_1 \mu_2 < 0$ the eigenvalues are not in a Poincar\'e domain and we are not guaranteed of the convergence of the normalizing transformation on the basis of the Poincar\'e criterion. If $\mu_1 / \mu_2$ is irrational, the NF is linear, and no further normalization is needed; moreover, convergence can be guaranteed on the basis of Pliss theorem \cite{Pli}. 
If $\mu_1 / \mu_2 = p/q \in {\bf Q}$, there can be resonances; in this case Sternberg theorem \cite{Arn2,Bel,BeK,Che,Ste} guarantees that the NF is smoothly (but in general, not analytically) equivalent to the original system. 

The PRF for this case has not been studied so far, and will be considered below in section 8.

In case {\bf N1} the eigenvalues are in a Poincar\'e domain, as $\mu \not= 0$;  moreover there are no resonances, and thus we have a linear NF, with a convergent normalizing transformation.

Finally, in the nonregular case {\bf N2} the standard normal form is given by $$ \begin{array}{rl} {\dot x} =& \sum_{k=1}^\infty a_k x^{k+1}  \\   {\dot y} =& \sum_{k=1}^\infty a_k x^k y + b_k x^{k+1} \ . \end{array} \eqno(\sn.2) $$ 
As already mentioned, the PRF for this case was studied in detail in \cite{IHP}. We briefly recall the (poor) results concerning this in section 10.

\bigskip

It follows from the above summary of known results that we need to study PRFs only in the cases {\bf S3} and {\bf S4} (and, as mentioned in remark 1, to correct a formula in case {\bf S2}). Actually, as discussed in the previous section, formal computations in one of these cases can be mapped to other cases as well. We will thus study the case {\bf S3} in full detail.

\section{The S3 case: standard normal forms}
\def\sn{5}

Let us consider a linear part given by 
$$ A  =  \pmatrix{0&0\cr0&1\cr} \eqno(\sn.1)$$
i.e. corresponding to the vector field
$ y \pa_y $. 
We note immediately that here $A = A^+$, so that the homological operator associated to $f_0 = Ax$ will satisfy $\L_0 = \L_0^+$ (recall we are using the Bargmann scalar product).   
It is easy to see that the kernel of $\L_0$ is spanned by the arrays of vector fields (with $k \ge 0$) 
$$ X_k  :=  x^{k+1} \, \pa_x \ \in \W_k \ \ {\rm and} \
\  Y_k  :=  x^k y \, \pa_y \ \in \W_k  \eqno(\sn.2) $$
(with this notation the linear part considered here is given by $Y_0$). These vector fields satisfy the commutation relations
$$  
[ X_k , X_m ] =   (m-k) \, X_{k+m} \ 
, \ \ 
[ X_k , Y_m ] =   m \, Y_m \
, \ 
\ 
[ Y_k , Y_m ] =   0 \ .  \eqno(\sn.3) $$
We denote by $\G$ the (infinite dimensional) Lie algebra spanned by the $X_k$'s and the $Y_k$'s; and by $\X$ the algebra spanned by the $X_k's$, by $\Y$ the
algebra spanned by the $Y_k$'s. 
Obviously $\G = \X \oplus \Y$. Note that $\Y$ is an abelian ideal,
actually the maximal abelian ideal, in $G$. 
The (standard) normal form corresponding to the linear 
part considered in this section will thus be given by a vector field
$$ W  =  Y_0  +  \sum_{k=1}^\infty ( a_k X_k + b_k Y_k ) \eqno(\sn.4) $$
depending on the two infinite sequences of real constants $a_m , b_m$.
This is precisely the structure considered in section 3.

{\bf Remark 9.} The vector fields $Z_- := \pa_x$, $Z_0 := x \pa_x$ and $Z_+ := x^2 \pa_x$ act in each of $\X$ and $\Y$, as respectively a lowering operator, a counting one, and a raising one: that is, $[Z_-,X_n]  =  (n+1) X_{n-1}$, $[Z_-,Y_n]  =  n Y_{n-1}$;   $[Z_+,X_n]  =  (n+1) X_{n+1}$,
$[Z_+,Y_n]  =  n Y_{n+1}$;   $[Z_0,X_n]  =  (n+1) X_{n}$, $[Z_0,Y_n]  =  n Y_{n}$. $\odot$

\section{The S3 case: Poincar\'e renormalized forms}
\def\sn{6}

We want now to consider the PRF corresponding to the linear part given by (5.1). In the spirit of PRF, we should act on the NF (5.4) with 
Lie-Poincar\'e transformations generated by homogeneous functions $h_m \in \ker (\L_0) \cap V_m$. These will correspond to the action of vector fields of the form $ H_m =  \alpha X_m  +  \beta Y_m$. 

We will first, in this section, consider the spaces defined in the PRF
procedure, and thus obtain the general form of the PRF in this case  (in the next section, we will perform explicit computations up to order $N=5$).
We recall that $\ker (\L_0 )$ corresponds to the sum of the algebras $\X$ and $\Y$, and that here $\ker (\L_0 ) = \ker ( \L_0^+ )$.

We have then to consider $\L_1$; this depends on the coefficients of the quadratic part $W_1$ of the vector field $W$, which we write as 
$$ W_1  =  a_1 X_1 + b_1 Y_1 \ . \eqno(\sn.1) $$
The cases to be considered are
$$ \cases{ 
a_1 \not= 0
\, , \, b_1 = 0 \, ; & (a) \cr 
a_1 = 0 \, , \, b_1 \not= 0 \, ; & (b) \cr 
a_1
\not= 0 \, , \, b_1 \not= 0 \, ; & (c) \cr 
a_1 = 0 \, , \, b_1 = 0 \, . & (d) \cr }
\eqno(\sn.2) $$
We refer to cases {\bf (a)},{\bf (b)},{\bf (c)} as nondegenerate
(although properly speaking only {\bf (c)} is such), and to {\bf (d)} as the degenerate (properly speaking, completely degenerate) case.

\subsection{The nondegenerate cases}

In case {\bf (a)} we have $W_1 = a_1 X_1$; we notice that
$$ [X_1 , X_k ] = (k-1) X_{k+1}  \ \ , \ \  [X_1 , Y_k ] = k
Y_{k+1} \eqno(\sn.3) $$
and therefore for $\M_1$ -- the restriction of $\L_1$
to $\ker (\L_0)$ -- we have that $ \ker (\M_1)$ reduces to the linear span of $\{ Y_0 ,  X_1 \}$, i.e. of $W_0$ and $W_1$, so no further normalization employing $\L_2, \L_3,...$ is possible.
We also have that the range of $\M_1$ (the most relevant space for our
discussion) is the whole linear span of the
$\{ X_k \}$ (with $k>2$) and of the $\{ Y_ k \}$ (with $k \ge 2$). 

{\bf Remark 10.} \def\rna{10}
We stress that for the sake of the present computations (which aim at identifying linear subspaces) we can
as well assume $a_1 =1$; the same remark would apply to other cases. Such a trivial remark will be of use later on in section 8. $\odot$
\bigskip

In case {\bf (b)} we have $W_1 = b_1 Y_1$.
We notice that 
$$ [Y_1 , X_k ] = - Y_{k+1}  \ , \  [Y_1 , Y_k ] = 0 \ , \eqno(\sn.4) $$
and therefore $\ker (\M_1 ) = \Y$. On the other hand, 
$\ran (\M_1)$ also is given by $\Y$, and $\ker (\M_1^+ ) = \X $.

In this case we also have to consider higher order parts of $W$; the first step of the PRF procedure can eliminate all terms in $\ran (\M_1 )$ and thus we will only consider terms in $\ker (\M_1^+)$. Let $p$ be the first integer for which $a_p \not= 0$, and let $W_p = a_p X_p$ (all the $Y_k$ parts with $k \ge 2$ can be eliminated, as just recalled). Now $\M_p$ is the restriction of
$\L_p$ to $\ker (\M_1 ) = \ker (\L_0 ) \cap \ker (\L_1)$: indeed the $\L_m$ with $1 < m < p$ are zero and put no restriction. We have
 $$ [X_p , Y_k ] = k Y_{k+p} \eqno(\sn.5) $$
 and thus $\ker (\M_p ) = \{ 0 \} $: no further
normalization is possible.
 
 In case {\bf (c)} the situation is quite similar to the one met in case {\bf (a)}: we have indeed $W_1 = a_1 X_1 +
b_1  Y_1 $ with nonzero constants $a_1 , b_1$; we have immediately that
$$ [ W_1 , \a X_k + \b Y_k ]  \ = \   a_1 \a (k-1)
X_{k+1} \, + \, (k a_1 \b - b_1 \a ) Y_{k+1} \eqno(\sn.6) $$
This shows that $\ker (\M_1)$ is just given by $\{ Y_0 , a_1 X_1 + b_1 Y_1 \}$, i.e. by the linear span of $W_0$ and $W_1$: so again no further normalization using operators $\L_2, \L_3, ...$ is possible. 
As for $\ran (\M_1)$, this is the linear span of $\{ X_k
\}$ with $k >2$, and of $\{ Y_k \}$ with $k \ge 2$.

\bigskip

We summarize the results of this discussion as follows, with $\^W$ the vector field after the whole PRF procedure and omitting the degenerate case {\bf (d)}. The hat on constants $\^a_k$ will indicate that coefficients are not the same as those of the initial NF (5.4).
 
$$ \^W  =  \cases{
 Y_0 + a_1 X_1 + \^a_2 X_2 & (a) \cr
 Y_0 + b_1 Y_1 +
\sum_{k=2}^\infty \^a_k X_k & (b) \cr
 Y_0 + a_1 X_1 + b_1 Y_1 + \^a_2 X_2 &
(c) \cr } \eqno(\sn.7) $$
 
We anticipate that a different reduction scheme (see section 3) can give a finite dimensional NF in case {\bf (b)}; we discuss this later on in subsection \sn.3.

\subsection{The degenerate case}

The discussion of the degenerate case {\bf (d)} would require to
consider the first $q$ such that $a_q^2 + b_q^2 \not=0$, and then repeat the considerations presented above in cases {\bf (a),(b),(c)}, obviously with the role of $a_1,b_1$ taken by $a_q,b_q$. This is done in the following lines.

We denote by $\mu > 1$ the first $k$ such that $a_k \not= 0$, and by $\nu > 1$ the first $k$ such that $b_k \not= 0$; at least one of these has to exist and be finite, or the system would already be linear and thus trivial.

We will have to consider three cases
 $$ \cases{
\mu < \nu & (da) \cr
\mu > \nu & (db) \cr
\mu = \nu & (dc) \cr} \eqno(\sn.8) $$

\bigskip

In case {\bf (da)} the NF will be given by 
$$ W \ = \ Y_0 \ + \ \sum_{k=\mu}^{\nu-1} a_k X_\mu \ + \
\sum_{k=\nu}^\infty (a_k X_k + b_k Y_k ) \ . \eqno(\sn.9) $$
 
We write again a $H_k \in \ker (\L_0 ) \cap \W_k$ as 
$H_k = \a_k X_k + \b_k Y_k$, and we have
$$ \begin{array}{rl}
\L_\mu (H_\nu ) \ := & \ \[ W_\mu , H_k \] \ = \ a_\mu \[
X_\mu , \a_k X_k + \b_k Y_k \] \ = \\  & \ = \ a_\mu \a (k-\mu ) X_{\mu+k} \
+ \ a_\mu \b_k k Y_{\mu+k} \ . \end{array} \eqno(\sn.10) $$

Thus it suffices to operate successively transformations generated by  $H_k$
(with $k=1,2,...$) and choose at each step
 $$ \a_k \ = \ {\=a_{\mu+k} \over
(k-\mu) a_\mu } \ \ , \ \ \b_k \ = \ {\=b_{\mu+k} \over k a_\mu} \ ,
\eqno(\sn.11) $$
 where $\=a_{\mu+k}, \=b_{\mu+k}$ denote the coefficient of
$X_{\mu+k}, Y_{mu+k}$ in $\=W$, i.e. after the action of previous
transformations.
Notice that in this way we can eliminate all terms except
the $X_{2 \mu}$ one ($k = \mu$). Thus, the PRF in case {\bf (da)} results to be
$$ \^W \ = \ Y_0 \ + \ a_\mu \, X_mu \ + \ \eta \, X_{2\mu} \eqno(\sn.12) $$
where $a_\mu$ is the same as in the NF and $\eta$ is a real number.
 
In case {\bf (db)} the NF is 
$$ W \ = \ Y_0 \ + \ \sum_{k=\nu}^{\mu-1} b_k Y_\mu \ + \
\sum_{k=\nu}^\infty (a_k X_k + b_k Y_k ) \ . \eqno(\sn.13) $$ 
Now we have for $\L_\nu (H_k)$ that
$$ \L_\nu (H_k) \ = \ b_\nu \[ Y_\nu , \a_k X_k + \b_k Y_k \] \ = \ - \nu b_\nu \a_k \, Y_{\nu+k} \eqno(\sn.14) $$
and therefore we can eliminate all the $Y_{\nu+k}$
terms simply by choosing, with the same notation as before,
$$ \a_k \ = \ { -
\=b_{\nu+k} \over \nu b_\nu} \ ; \eqno(\sn.15) $$
we cannot eliminate any of the $X_k$ terms. 
 
Thus, the PRF in case {\bf (db)} is
$$ \^W \ = \ Y_0 \ + \ b_\nu Y_\nu \ + \ \sum_{k=\mu}^\infty \=a_k X_k \ . \eqno(\sn.16) $$
 
Similarly to what happens for the nondegenerate case {\bf (b)}, the  reduction scheme based on LRF, discussed below, gives better results in this case.
 
In case {\bf (dc)} we have $\mu=\nu$; the NF is
$$ W \ = \ Y_0 \ + \ \sum_{k=\mu}^\infty (a_k X_k + b_k Y_k ) \ . \eqno(\sn.17) $$
In  this case
$$ \begin{array}{rl}
\L_\mu (H_k) \ = & \ \[a_\mu X_\mu + b_\mu Y_\mu , \a_k X_k + \b_k Y_k \] \ =
\\ & = \  a_\mu \a_k (k-\mu) X_{\mu+k} \ + \ (k a_\mu \b_k - \mu b_\mu \a_k )
Y_{\mu+k} \ . \end{array} \eqno(\sn.18) $$
 Thus for $k \not= \mu$ it
suffices to choose
 $$ \a_k \ = \ {\=a_{\mu+k} \over (k-\mu) a_\mu} \ \ , \ \
\b_k \ = \ { (k-\mu) a_\mu \=b_{\mu+k} + \mu b_\mu \=a_{\mu+k} \over a_\mu^2 k
(k-\mu) } \eqno(\sn.19) $$
 to eliminate both the $X_{\mu+k}$ and the
$Y_{\mu+k}$ terms. 
For $k = \mu$, we choose $\a_k = 0$ and $\b_k = \=b_{\mu+k}
/ (k a_\mu)$ and eliminate the $Y_{2 \mu}$ term. 
 Thus, the PRF in case {\bf (dc)} is
$$ \^W \ = \ Y_0 \ + \ a_\mu X_\mu \ + \ b_\mu Y_\mu \ + \ \eta X_{2 \mu} \ . \eqno(\sn.20) $$
 
\subsection{LRF reduction scheme for cases (b) and (db)}

In the previous computations, we have followed the general PRF scheme for further normalizing the standard NF (5.4); this gave an infinite PRF in case {\bf (b)} and in the corresponding degenerate case {\bf (db)}. 
However, as discussed in subsection 3.3, one can take advantage of the specific Lie algebraic structure of $\G_A = \X \oplus_\to \Y$ (this recalls that $\X$ is acting on $\Y$ by inner automorphisms) to obtain a more drastical reduction: indeed, in this case the Lie renormalized form will be a finite one, as we now discuss. We use the same notation as in discussing the case {\bf (db)} above.
 
We first operate a sequence of normalizations with
generators $h_k^{(a)} = \a_k X_k$, which we choose so as to eliminate higher order $X_k$ terms, i.e. $X_k$ for $k > \mu$ (as we know, this is not possible for $k = 2 \mu$). Notice this will change not only the (coefficients of the) $X_k$
terms, but the (coefficients of the) $Y_k$ terms as well; however, no terms of degree $k < \nu$ will be produced. 
 In this way, we arrive at a form of the
type (as usual the tilde indicates that the coefficients are not the same as the initial ones, but not yet final)
 $$ \=W \ = \ Y_0 \ + \ a_\mu X_\mu + \=a_{2 \mu} X_{2 \mu}
\ + \  \sum_{k=\nu}^\infty \=b_k Y_k \ . \eqno(\sn.21) $$

Once this has been done, we pass to consider a second sequence of normalizations with generators $h_k^{(b)} = \b_k Y_k $. As $\Y$ is an ideal in $\G_A$, in this way only $Y_k$ terms are generated, i.e. the $X_k$ terms are unaffected. 
On the other side, $\Y$ is abelian, and so only the $X_\mu$ and $X_{2 \mu}$ are actually active in these transformations: that is, we can only eliminate terms $Y_{\mu + 1}$ and higher (it is clear by the commutation relations that these can always be eliminated). 
 
In this way, we arrive at the LRF: this is a NF depending on $(\mu - \nu + 3)$ constants (this agrees with the number of constants predicted by Bruno in sect. III.2.3 of \cite{Bru}; see also \cite{Brus}), of the form
 $$ \^W \ = \ Y_0  \ + \ a_\mu X_\mu + \^a_{2 \mu} X_{2 \mu} \ + \ 
\sum_{k=\nu}^\mu \^b_k Y_k \ . \eqno(\sn.22) $$
It is also clear by this discussion that $\^b_k = \=b_k$, $\^a_{2 \mu } = \=a_{2 \mu}$.
 
{\bf Remark 11.} It should be stressed that the LRF (22) is {\it not } a PRF. Indeed, in this case the spaces $F^{(k)}_k := F^{(k)} \cap \W_k$ with $\nu < k \le \mu$ reduce, as seen in subsection \sn.2, to multiples of $X_k$. Here we have therefore $W_k \not\in F^{(k)}_k$ for $\nu < k \le \mu$, and thus (see section 2) the LRF cannot be a PRF.
 $\odot$

\section{The S3 case: explicit computations}
\def\sn{7}

As stressed in \cite{LMP,IHP}, the PRF procedure is completely constructive; indeed, the PRF procedure gives an algorithm (which is easy to implement on a computer, as shown by the formulas reported in this section) to determine the coefficients $\a_k , \b_k$ of transformations needed to take the system (5.4) into its PRF.
   
In this section I follow these computations (not needed if one is only interested in the most general PRF form) for the linear part $A$ given by (5.1), up to order six in $x$ and $y$ -- i.e. put $W$ in PRF up to terms $W_5$ -- for the nondegenerate cases.\footnote{Explicit formulae for a specific degenerate case which raised some confusion in the literature are given in the appendix, up to order ten in $x$ and $y$.}
To avoid trivial steps, I will always assume that a first Poincar\'e normalization has already been performed, taking the system into its standard normal form $f^{(1)}$ ($W = W^{(1)}$). 
 
In order to display the rather long explicit formula we obtain, it will be convenient to denote by $W_k$ the part of the (coordinate expression of the) vector field $W$ homogeneous of degree $k+1$ in the coordinate we are using; note they will change with the changes of coordinates.
 
Recall the explicit expressions given below are computed using the Baker-Campbell-Haussdorff formula (1.7), (1.9); i.e. by considering Lie-Poincar\'e transformations, and not simply Poincar\'e ones.
   
\subsection{Case (a)}
 
We first operate a transformation with $h_1 \in \ker (\L_0) \cap \F_1$, i.e. with  $H_1 = \a_1 X_1 + \b_1 Y_1 \in \W_1
\cap \ker (\L_0)$; after this, the quadratic part $W_1$ of the vector field is
unchanged, $W_1^{(2)} = W_1^{(1)}$, while the cubic one is given by 
 $$ W_2^{(2)} \ = \ a_2 X_2 \, + \, (b_2 - a_1 \b_1) Y_2  \ . \eqno(\sn.1)$$
We know from our previous general discussion that -- as indeed obvious from the above formula -- the first component of this cannot be eliminated; to eliminate the second one, we have to choose 
$$ \b_1 \ = \ b_2 / a_1 \ ; \eqno(\sn.2)$$ 
we can choose $\a_1$ as we like, say 
$$ \a_1 \ = \ 0 \eqno(\sn.3)$$ 
for simplicity. This choice of $\a_1,\b_1$ fixes the PRF after
the first renormalization, i.e. $f_k^{(2)}$. We do not give the explicit formulae for the sake of brevity.
  
Let us now operate a transformation with $h_2  \in \F_2 \cap \ker (\L_0)$; the terms $f_0,f_1,f_2$ are unaffected. Using the explicit formulae for $f_k^{(2)}$, we have that $W_3^{(2)}$ is changed into
 $$ W_3^{(3)} \ = \
 (a_3 - a_1 \a_2) \, X_3 \ + \ 
 (b_3 - 2 a_1 \b_2 - a_2 b_2/a_1 ) \, Y_3  \ . \eqno(\sn.4)$$
We know from our previous discussion that we should be able to eliminate both components of this vector; this can indeed be obtained by choosing 
 $$ 
\a_2 \, = \, \frac{a_3}{a_1} \ \ , \ \ 
\b_2 \, = \, \frac{\left( a_1 b_3 - a_2 b_2 \right) }{2
a_1^2} \eqno(\sn.5)$$
This choice of $\a_2,\b_2$ fixes the PRF after the second renormalization, i.e. the $f_k^{(3)}$. Again we do not give the explicit formulae.
 
Let us now operate a transformation with $h_3  \in \F_3 \cap \ker (\L_0)$; the terms $f_0,...,f_3$ are unaffected. Using the explicit formulae for $f_k^{(2)}$ and $f_k^{(3)}$, we have that $W_4^{(3)}$ is changed into
$$ \begin{array}{rl}
W_4^{(4)} \ =& \ \left( a_4 - 2 a_1 \a_3 \right) \ X_4 \ + \\
 & \ + \  \left( \left[  a_2^2 b_2 - a_1 a_2 b_3 + a_1 \left( - a_3 b_2 + a_1
 \left( b_4 - 3 a_1 \b_3
\right)  \right)  \right] / (a_1^2) \right) \ Y_4 \end{array} \eqno(\sn.6)$$
Again we know apriori that this can be eliminated, and indeed the above formula shows that this is the case if we choose
$$ \begin{array}{ll}  
\a_3 =& \ a_4/(2 a_1) \\
\b_3 =& \ [  a_2^2 b_2 -  a_1 a_3 b_2 - a_1 a_2 b_3 + a_1^2 b_4 ] / (3 a_1^3) \end{array} \eqno(\sn.7)$$
This choice of $\a_3,\b_3$ fixes $f_k^{(4)}$. 
 
Let us now operate a transformation with $h_4  \in \F_4 \cap \ker (\L_0)$; the terms $f_0,...,f_4$ are unaffected. Using the explicit formulae for $f_k^{(2)}$, $f_k^{(3)}$ and $f_k^{(4)}$, we have that $W_5^{(4)}$ is changed into
$$ \begin{array}{ll}
W_5^{(5)} \ =& \ 
[ ( a_3^2 - a_2 a_4 + 2 a_1 ( a_5 - 3 a_1
\a_4 ) ) \,  / \, (2 a_1) ] \ X_5 \\ 
 & + \ [ ( a_1^2
( - a_4 b_2 + a_3 b_3 - a_2 b_b + a_1 b_5 - 4 a_1^2 \b_4 ) \\
 & \ \ - a_2^3 b_2 + a_1 a_2^2 b_3 ) \, / \, (a_1^3) ] \ Y_5  \end{array}
\eqno(\sn.8)$$
 which goes to zero if we choose
$$ \begin{array}{ll}
\a_4\ =& \ \left( a_3^2 - a_2 a_4 + 2 a_1 a_5 \right) \, / \, ( 6 a_1^2 ) \\
\b_4\ =& \ - \left( a_2^3 b_2 + a_1^2 a_4 b_2 -
a_1 a_2^2 b_3 - a_1^2 a_3 b_3 + a_1^2 a_2 b_4 - 
a_1^3 b_5 \right) \, / \, (4 a_1^4) \ . \end{array} \eqno(\sn.9)$$
 
Clearly, the computation could be performed up to any desired order,
compatibly with the computational power at our disposal, producing more and more complex but still completely explicit formulae; we will stop at this order.

\subsection{Case (b)}

Let us now consider the (slightly more complex) case {\bf (b)}.
This was subject to some controversy (see remark 2 and the appendix), so that we will discuss it in full detail, following the transformation of the coordinate expression of the vector field $W$ step by step. 

With a first transformation generated by $H_1 = \a_1 X_1 + \b_1 Y_1$ we have that $\~W_2^{(2)}$ is given by $a_2 X_2 + (b_2 + \a_1 b_1) Y_2$; requiring the coefficient of $Y_2$ to vanish, we get
$$ \a_1 \ = \ - b_2 / b_1 \ , \eqno(\sn.10)$$ 
and for the sake of simplicity we will take $\b_1 = 0$. 
In this way we get (here and in the following $O(6)$ denote terms in $\W_6$ and higher)
$$ \begin{array}{rl}
\~W^{(2)} \ = &\   
 Y_0 \ + \ b_1 \, Y_1 \ + \ a_2 \, X_2 \ + \\
 & + \ [ a_3 - a_2 b_2 / b_1 ] \, X_3 \ 
 + \ [ b_3 - b_2^2 / b_1 ] \, Y_3 \ + \\
 &  + \ [ a_4 - 2 a_3 b_2 / b_1 +  a_2 b_2^2 / b_1^2  ] \, X_4 \ + \\
 & + \ [ b_4 + 2 b_2^3 / b_1^2 - 3 b_2 b_3 / b_1 ] \, Y_4 + \\
 & + \ [ a_5 - 3 a_4 b_2 / b_1 + 3 a_3
b_2^2 / b_1^2 - a_2 b_2^3 / b_1^3 ] \, X_5 + \\
  & + \ [ b_5 -3 b_2^4/b_1^3 + 6
b_2^2 b_3/b_1^2 - 4 b_2 b_4 /b_1 ] \,  Y_5 \\ & + \ O(6) \ . \end{array} \eqno(\sn.11)$$
 
We will now operate a transformation generated by $H_2 = \a_2 X_2 + \b_2 Y_2$. This leaves lower order terms unaffected, while $W_3$ reads after this 
$$ \~W_3^{(3)}  =  
 ( a_3 - (x^4 a_2 b_2)/b_1 ) X_3 +  
( (-((x^3 y b_2^2)/b_1)) + x^3 y b_3 + 
      x^3 y b_1 \a_2) ) Y_3 \eqno(\sn.12)$$
Requiring the vanishing of the coefficient of the $Y_3$ term we get 
$$ \a_2  \ = \ (b_2^2 - b_1 b_3) \, / \, (b_1^2) \ ; \eqno(\sn.13)$$ 
we will set again $\b_2 = 0$ for the sake of simplicity.
With these, we get
$$ \begin{array}{rl}
 \~W^{(3)} \ = & \  
Y_0 \ + \ b_1 \, Y_1 \ + \ a_2 \, X_2 \ + \ 
[ a_3 - a_2 b_2/b_1 ] \, X_3 \ + \\ 
 &  + \ [ a_4 - 2 a_3 b_2 / b_1 + a_2 b_2^2 / b_1^2 ] \, X_4 \ + \\  
 &  + \ [ 2 b_2^3 / b_1^2 - 3 b_2 b_3 / b_1 + b_4 ] \, Y_4 \ + \\ 
 &  + \ [ a_5 - 3 a_4 b_2 / b_1 + 3 a_3 b_2^2 / b_1^2 - a_2 b_2^3 / b_1^3 + \\
 & \ \ \ \ + \ (a_3 b_1 - a_2 b_2) (b_2^2 - b_1 b_3) / b_1^3  ] \, X_5 \ + \\ 
 &  + \ [ 9 b_2^4 / b_1^3 + 9 b_2^2 b_3 / b_1^2 - 3 b_3^2 / b_1 - 4 b_2 b_4
/ b_1 + b_5 ] \, Y_5 \ + \ O(6) \ . \end{array} \eqno(\sn.14)$$

Let us now consider a transformation with generator $H_3 = \a_3 X_3 + \b_3 Y_3$. Now lower order terms are unaffected, while we get that the coefficient of the $Y_4$ term is changed to 
 $$ (2 b_2^3)/(b_1^2) \, - \, (3 b_2
b_3)/(b_1) \, + \,  
 b_4 \, + \, b_1 \a_3 \ . \eqno(\sn.15)$$
Requiring this to vanish, we get 
$$ \a_3  \ = \  - \, (2 b_2^3 - 3 b_1 b_2 b_3 + b_1^2 b_4) \, / \, (b_1^3) \ .
\eqno(\sn.16)$$
We will, as by now usual, set $\b_3 = 0$; with these we
obtain 
$$ \begin{array}{rl}
~W^{(4)} \ = &\   
Y_0 \ + \ b_1 \, Y_1 \ + \ a_2 \, X_2 \ + \ 
[ a_3 - a_2 b_2/b_1 ] \, X_3 \ + \\
 & + \ [ a_4 - 2 a_3 b_2 / b_1 + a_2 b_2^2 / b_1^2 ] \, X_4 \ + \\
 & + \ [ a_5 - 3 a_4 b_2 / b_1 + 4 a_3 b_2^2 / b_1^2 - 
a_3 b_3 / b_1 \ + \\
 & \ \ \ \ - 2 a_2 b_2 b_3 / b_1^2 + 
a_2 b_4 / b_1 ] \, X_5 \ + \\
 & + \ [ - 9 b_2^4 /(2 b_1^3) + 9 b_2^2 b_3 / b_1^2 \ + \\
 & \ \ \ \ - 3 b_3^2 / (2 b_1) - 4 b_2 b_4 / b_1 + b_5) ] \, Y_5 \ + \ O(6)
\ .  \end{array} \eqno(\sn.17)$$
  
We now operate with $H_4 = \a_4 X_4 + \b_4 Y_4 $; the coefficient of $Y_5$ results to be
$$ -(9 b_2^4)/(2 b_1^3) \, + \, (9 b_2^2 b_3)/(b_1^2 ) \, - \,  
(3  b_3^2)/(2 b_1) \, - \, (4 b_2 b_4)/(b_1) \, + \, b_5 \, + \, b_1 \a_4 \ . \eqno(\sn.18)$$
Requiring this to vanish, we get 
$$ \a_4  \ = \ (9 b_2^4 - 18 b_1 b_2^2 b_3 + 3 b_1^2 b_3^2 + 
8 b_1^2 b_2 b_4 - 2 b_1^3 b_5) \, / \, (2 b_1^4) \ ; \eqno(\sn.19)$$ we also set $\b_4 = 0 $.
We now have
$$ \begin{array}{rl}
\~W^{(5)} \ = &\ Y_0 \ + \ b_1 \, Y_1 \ + \ a_2 \, X_2 \ + \ 
[ a_3 - a_2 b_2 / b_1 ] \, X_3 \ + \\
 & + \ [ a_4 - 2 a_3 b_2 / b_1 + a_2 b_2^2 / b_1^2 ] \, X_4 \ + \\
 & + \ [ a_5 - 3 a_4 b_2 / b_1 + 4 a_3 b_2^2 / b_1^2 + \\
 & \ \ \ \ - a_3 b_3 / b_1 - 2 a_2 b_2 b_3 / b_1^2 + a_2 b_4 / b_1 ] \, X_5
\ + \ O(6) \ . \end{array} \eqno(\sn.20)$$
Again we will stop at this order; the result of this explicit computation fits in the general result obtained in the previous section.

\subsection{Case (c)}

We could analyze the other case {\bf (c)} and produce explict formulas
proceeding in the same way as in the previously considered cases {\bf (a)} and {\bf (b)}; however, the procedure is by now clear and for the sake of brevity we will just give the final formulae. 
  
The coefficients $\a$ are chosen as
$$ \begin{array}{rl}
\a_1 \ =\ & 0 \\
\a_2 \ =\ & {a_3}/{a_1} \\ 
\a_3 \ =\ & {a_4}/(2 a_1) \\
\a_4 \ =\ & \left( a_3^2 - a_2 a_4 + 2 a_1 a_5 \right) \, / \, ( 6 a_1^2) \ ;
\end{array} \eqno(\sn.21)$$
the coefficients $\b$ are chosen as
$$ \begin{array}{rl}
\b_1  \ =\ & b_2 \, / \, a_1 \\ 
\b_2  \ =\ & \left( a_3 b_1 - a_2 b_2 + a_1 b_3 \right) \, / \, ( 2 a_1^2) \\
\b_3  \ =\ & -\left( 2 a_2 a_3 b_1 - a_1 a_4 b_1 - 2 a_2^2 b_2 + 2 a_1 a_3 b_2 + 2 a_1 a_2 b_3 - 2 a_1^2 b_4 \right) \, / \, ( 6 a_1^3) \\ 
\b_4  \ =\ & ( 3 a_2^2 a_3 b_1 - a_1 a_3^2 b_1 - 2 a_1 a_2 a_4 b_1 + a_1^2 a_5 b_1 - 3 a_2^3 b_2 - 3 a_1^2 a_4 b_2  + \\
 & \ \ \ + 3 a_1 a_2^2 b_3 + 3 a_1^2 a_3 b_3
- 3 a_1^2 a_2 b_4 + 3 a_1^3 b_5 ) \, / \, ( 12 a_1^4 ) \ . \end{array}
\eqno(\sn.22)$$
  
It this way, we arrive at a PRF given by 
$$ \=W^{(5)} \ = \ Y_0 \, + \, a_1 X_1 \, + \, b_1 Y_1 \, + \, 
 a_2 X_2 \, + \, O(6) \ . 
\eqno(\sn.23)$$
 This corresponds to the result obtained by our general
discussion above (moreover the coefficient of the $X_2$ term is
unchanged).

\subsection{The LRF scheme for case (b)}

As mentioned in subsection 6.3, in case {\bf (b)} the alternative scheme adapted to the structure of $\G$ described there permits to obtain a finite NF (the LRF) and is thus to be preferred to the general one. Here we briefly discuss the explicit computation to be performed according to this. We deal with the nondegenerate (properly speaking, not completely degenerate) case, which means $\mu= 2$, $\nu=1$; see subsection 6.3.
 
With a transformation $h_1 = \a_1 X_1$, the $W_3$ term reads
$$ \=W_3 \ = \ \[ a_3 + a_2 \a_1 \] X_3  \ + \ \[ b_3 + 2 b_2 \a_1 + b_1 \a_1^2 \] Y_3 \ . \eqno(\sn.24) $$
We disregard the $Y_3$ term and choose $\a_1$ so to eliminate the $X_3$ term, i.e. $\a_1 = - a_3 / a_2$. 
After computing the effect of this on higher order terms, we could perform a transformation with generator $h_2 = \a_2 X_2$. However, we know that there will be no way to eliminate the $X_4$ term, so we set $\a_2 = 0$. We perform a transformation with generator $h_3 = \a_3 X_3$.  With this, the $W_5$ term reads
$$ \begin{array}{rl}
\=W_5 \ =& \ [ 2 a_3^3 /a_2^2 - 3 a_3 a_4 / a_2 + a_5 - a_2 \a_3 ] X_5 \ + \\ 
 & + \ [ a_3^4 b_1 / a_2^4 - 4 a_3^3 b_2 / a_2^3 + 
   6 a_3^2 b_3 / a_2^2 - 4 a_3 b_4 / a_2 + \\ 
 & + \ \ \ b_5 - 2 a_3 b_1 \a_3 / a_2 + 2 b_2 \a_3 ] Y_5 \ . 
\end{array} \eqno(\sn.25)  $$
Again we only aim at eliminating the $X_5$ term, and thus we choose
$$ \a_3 \ = \ { 2 a_3^3 - 3 a_2 a_3 a_4 + a_2^2\ a_5 \over a_2^3} \ . \eqno(\sn.26) $$
We will be satisfied with this order of normalization for the $X_k$ terms, and take now care of the $Y_k$ ones.
 
We operate a transformation with generator $h_1 = \b_1 Y_1$; with this we have that 
$$ \=W_3 \ = \ \[ a_3^2 b_1 / a_2^2 - 2 a_3 b_2 / a_2 + b_3 -  a_2
\b_1 \] \ Y_3 \ . \eqno(\sn.27) $$
We eliminate this by choosing 
$$ \b_1 \ = \ { a_3^2 b_1 - 2 a_2 a_3 b_2 + a_2^2 b_3 \over  a_2^3} \ . \eqno(\sn.28) $$
We compute the effect on higher order term, and then consider a transformation with generator $h_2 = \b_2 Y_2$; with these, we have
$$ \begin{array}{rl}
\=W_4 \ =& \ \[ a_4 - a_3^2 / a_2 \] \ X_4  \ + \\  
 & \ + \ (1/a_2^3) \ [ a_3^3 b_1 + 3 a_2 a_3^2 b_2 - 
        3 a_2 a_3 (a_4 b_1 + a_2 b_3) +  \\
 & \ \ +  a_2^2 (a_5 b_1 + a_2 (b_4 - 2 a_2 \b_2)) \, ] \ Y_4 \ . 
\end{array} \eqno(\sn.29)$$
We want to eliminate the $Y_4$ term, and thus we choose
$$ \b_2 \ = \ {1 \over  2 a_2^4} \  ( a_3^3 b_1 - 3 a_2 a_3 a_4 b_1 + a_2^2 a_5 b_1 + 
3 a_2 a_3^2 b_2 - 3 a_2^2 a_3 b_3 + a_2^3 b_4) \eqno(\sn.30)$$
Again we take into account the effect of this on higher order terms, and pass to consider a transformation with generator $h_4 = \b_4 Y_4$; we get 
$$ \begin{array}{rl}
\=W_5 \ =& \ [ - 2 a_3^4 b_1 / a_2^4 + 
   5 a_3^2 a_4 b_1 / a_2^3 -  2 a_3 a_5 b_1 / a_2^2 - 
   2 a_3^3 b_2 / a_2^3 - 4 a_3 a_4 b_2 / a_2^2 + \\
 & \ + 2 a_5 b_2 / a_2 + 7 a_3^2 b_3 / a_2^2 - 
   a_4 b_3 / a_2 - 4 a_3 b_4 / a_2 + b_5 - 3 a_2 \b_3 ] \ Y_5 
\end{array} \eqno(\sn.31)$$
which can be eliminated by choosing 
$$ \begin{array}{rl}
\b_3 \ =& \ 
- \ (1 /( 3 a_2^5 )) \ 
(2 a_3^4 b_1 - 5 a_2 a_3^2 a_4 b_1 + 
 2 a_2^2 a_3 a_5 b_1 + 2 a_2 a_3^3 b_2 + \\
 & \ + 4 a_2^2 a_3 a_4 b_2 - 2 a_2^3 a_5 b_2 - 
 7 a_2^2 a_3^2 b_3 + a_2^3 a_4 b_3 + 
 4 a_2^3 a_3 b_4 - a_2^4 b_5 ) \ . \end{array} \eqno(\sn.32)$$

Summarizing, and having taken into account all higher order effects (up to order six), we have reached the LRF
$$ \^W \ = \ Y_0  +  b_1 Y_1  +  a_2 X_2  +  [ b_2 - (a_3 b_1 / a_2 ) ] Y_2  +  [a_4 - (a_3^2 / a_2 ) ] X_4  + \ O(6) \eqno(\sn.33) $$
We will be satisfied with this order of normalization.

\subsection{Changes of coordinates}

In this section we have given completely explicit formulas for the generators of Lie-Poincar\'e transformations and for the PRF which can be obtained in this way for case {\bf S3} (these are also applied to other cases, as discussed in section 3 and also in sections 9 and 10).
It should be mentioned that in this simple case, one can describe exactly the change of coordinates generated by the vector field $H_k = \a_k X_k + \b_k Y_k$.

The evolution under the vector field $H_k$ ($k \ge 1$) is described by 
$$ {d x \over d s} = \a_k x^{k+1} \ \ , \ \ {d y \over ds} = \b_k x^k y \eqno(\sn.34)$$
with initial datum $x(0)= x_0 , y(0)=y_0$.
The first of these is solved by elementary methods to give
$$ x(s) \ = \ { x_0 \over ( 1 - \a_k k s x_0^k )^{1/k} } \ \ . \eqno(\sn.35)$$
Using this expression for $x(s)$, the second of (\sn.34) is rewritten 
$$ {dy \over y} \ = \ \b_k \ {x_0^k \over ( 1 - \a_k k s x_0^k )} \ ds \ , \eqno(\sn.36)$$
which gives
$$ y(s) \ = \ y_0 \ { 1 \over ( 1 - \a_k k s x_0^k )^{\b_k / (\a_k k)} } \eqno(\sn.37) $$

We are interested in the mapping $(x,y) = (x_0,y_0) \to (x(1),y(1) ) := (\=x , \=y )$, and from the above we have that
$$ \=x \ = \ { x \over ( 1 - \a_k k x^k )^{1/k} } \ \ , \ \ 
\=y \ = \ y \ { 1 \over ( 1 - \a_k k x^k )^{\b_k / (k \a_k)} } \eqno(\sn.38)$$
with the inverse change of coordinates given by
$$ x \ = \ {\=x \over 1 + \a_k \=x} \ \ ; \ \ 
y \ = \ \left( 1 - {\a_k \=x \over 1 + \a_k \=x } \right)^{\b_k / \a_k} \ \=y \ . \eqno(\sn.39)$$

These allow to obtain explicitely the changes of coordinates performed in passing from NFs to PRFs in case {\bf S3}, and can be mapped to consider the other cases as well. The explicit formulas, however, would contain rational power and be quite involved.

In order to make contact with the explicit formulas given above, notice that if we are acting with $H_k$, we are actually considering the change of coordinates from $x^{k}$ to $x^{(k+1)}$.
This map is defined only for $x^k < (\a_k k)^{- 1}$; this allows therefore to explicitely compute the domain of analyticity of the change of coordinates.

\section{The S4 case: standard normal forms}
\def\sn{8}

We consider now the case {\bf S4}, i.e. the linear part of our vector field is now given by 
$$ A \ = \ \pmatrix{ \la & 0 \cr 0 & \mu \cr } \eqno(\sn.1) $$
with $\la \not= \mu$, $ \la \not= 0$ and $\mu \not= 0$. 

As remarked in section 4, if $\la \mu > 0$ (the eigenvalues have the same sign), we are in a Poincar\'e domain, so the
convergence of the transformation to NF is guaranteed; on the other
hand, if $\la \mu < 0$ (i.e. we have an hyperbolic saddle point in the
origin), we are not in a Poincar\'e domain. However, the Chen-Sternberg theorem \cite{Arn2,Bel,BeK,Che,Ste} guarantees the system is $C^\infty$ conjugated to its normal form; as for the analytic conjugacy in this case, this is guaranteed if $|\la / \mu|$ is irrational, due to Pliss' theorem \cite{Pli}.

We also noticed that if $\la / \mu $ is irrational, there are no
resonances, i.e. the NF is linear; in this case we do not need (nor it makes sense) to consider PRFs.
Let us thus focus on the rational case. We will assume $|\la / \mu|  = p/q$, i.e. $ |\la| = c q$, $|\mu| = c p$, with $p$ and $q$ positive integers relatively prime.

\subsection{Eigenvalues having the same sign}
 
We should first of all notice that for $\la \mu > 0$, no resonances are actually possible unless one of the eigenvalues is a multiple of the other, and in this case we have only one resonant term. 
 
Indeed for $\la \mu > 0$ the only possible resonant terms are given by 
$$ \vb \ = \ \pmatrix{y^{\la / \mu}\cr 0\cr} \ (\la / \mu \in \N) \ \ {\rm or} \ \
\vb \ = \ \pmatrix{0\cr x^{\mu / \la}\cr} \ (\mu / \la \in \N) \ .
\eqno(\sn.2)$$
In order to see this, recall the resonance relations are now $m_1 \la + m_2 \mu = \la $ for the $x$ component, and $m_1 \la + m_2 \mu = \mu $ for the $y$ component. These give $(m_1 -1) \la + m_2 \mu = 0$ for the $x$ component, and $m_1 \la + (m_2 - 1) \mu = 0$ for the $y$ component; here $m_1 , m_2$ are non-negative integers, and $m_1 + m_2 \ge 2$. As $\la , \mu$ have the same sign, the only possibility in the $x$ case is $m_1 = 0$, and $ \la = m_2 \mu$ with $m_2 \ge 2$. Similarly, in the $y$ case it must be $m_2 = 0$, and thus $ \mu = m_1 \la $ with $m_1 \ge 2$.

We have thus shown that for $\la \mu > 0$ the standard NF for the case {\bf S4} is given, with $\a,\b$ are arbitrary real constants, by 
$$ \cases{ {\dot x} = \la x + \a y^k \ , \ {\dot y} = \mu y & for $\la/\mu = k \in \N$, $k \ge 2$ \cr
{\dot x} = \la x \ , \ {\dot y} = \mu y + \b x^k & for $\mu/\la = k
\in \N$, $k \ge 2$ \cr
{\dot x} = \la x \ , \ {\dot y} = \mu y & otherways \ . 
\cr} \eqno(\sn.3)$$
In each of these cases the PRF is trivial, i.e. it coincides with the
standard NF. We will thus give no further consideration to the case $\la \mu > 0$.
 
\subsection{Eigenvalues with opposite signs}
 
Consider now the rational case with $\la \mu < 0$. Assume
$$ \la = c q \ , \ \mu = - c p \eqno(\sn.4)$$
with $p,q$ positive integers, relatively prime (no common factor), and $c \not= 0$ a real number (notice we could have $p=q=1$, corresponding to $\mu = - \la$). For the sake of our discussion, we could as well  take $c = 1$.

The resonance relations give now $ (m_1 - 1) \la = - m_2 \mu$ for the $x$ component, and $m_1 \la = (1 - m_2) \mu$ for the $y$ component.
Hence we must have, for the $x$ component, $ m_2 / (m_1 -1) = - \la / \mu = q/p $, i.e.
$$ m_1 = k p + 1 \ , \ m_2 = k q \ . \eqno(\sn.5)$$
Similarly, for the $y$ component we have $ (m_2 -1)/ m_1 = - \la / \mu = q/p $, and therefore 
$$ m_1 = k p \ , \ m_2 = k q + 1 \ . \eqno(\sn.6)$$
Thus, the resonant vectors are of two types: 
$$ \vb_k^{(x)} \ = \pmatrix{(x^p y^q)^k \, x \cr 0 \cr}\ \ , \ \  
\vb_k^{(y)} \ = \ \pmatrix{0\cr (x^p y^q)^k \, y \cr} \ .  \eqno(\sn.7)$$
Correspondingly we consider vector fields
$$ \Phi_k \ = \ [(x^p y^q)^k \, x ] \, \pa_x \ \ , \ \ 
   \Psi_k  \ = \ [(x^p y^q)^k \, y ] \, \pa_y \ . \eqno(\sn.8)$$

The most general NF will be in the form $W = W_0 + \sum( c_k^{(1)} \Phi_k + c_k^{(2)} \Psi_k )$, i.e. 
$$ {\dot x} \ = \ \la x \, + \, \sum_{k=1}^\infty \, c_k^{(1)} (x^p y^q)^k \, x \ \ , \ \ 
{\dot y} \ = \ \mu y \, + \, \sum_{k=1}^\infty \, c_k^{(2)} (x^p
y^q)^k \, y \ .  \eqno(\sn.9)$$
The corresponding vector field will be denoted as $W$; its linear part is given by $W_0 = q \Phi_0 - p \Psi_0$.
 
For our discussion it will actually be more convenient to consider linear combinations of the $\Phi_k , \Psi_k$, defined as
$$ \begin{array}{rll}
X_k \ = \ & \( { 1 \over 2 pq }\) \ (q \Phi_k + p \Psi_k) \ = \ & {1 \over
2pq} \, (x^p y^q)^k \, ( q x \pa_x \, + \, p y \pa_y ) \ , \\
Y_k \ = \ & \( { 1 \over 2 pq }\) \ (q \Phi_k - p \Psi_k) \ = \ & 
{1 \over 2pq} \, (x^p y^q)^k \, ( q x \pa_x \, - \, p y \pa_y ) \ ;
\end{array}  \eqno(\sn.10)$$
with this notation, the linear part $W_0$ of the vector field $W$ corresponds to $W_0 = 2cpq Y_0 := \zeta Y_0$.
 
We also rewrite the corresponding vector field $W$, in view of the use of the vector fields $X_k$ and $Y_k$ and for further discussion, as 
$$ W \ = \ \zeta Y_0 \ + \ \sum_{k=1}^\infty ( a_k X_k + b_k Y_k )
\eqno(\sn.11)$$
where 
$$ a_k \ = \ (p c_k^{(1)} + q c_k^{(2)}) \ \ , \ \ 
b_k \ = \ (p c_k^{(1)} - q c_k^{(2)}) \ . \eqno(\sn.12)$$

{\bf Remark 12.} \def\rnn{12} Notice that $\Phi_k , \Psi_k \in \W_k$, but, with $z = p + q$, we have instead $X_k , Y_k \in \W_{kz}$. $\odot$

The vector fields $\Phi_k$ and $\Psi_k$ satisfy the commutation relations
$$ \begin{array}{lll}
 \[ \Phi_k , \Phi_m \] & \ =  \ & p (m-k) \, \Phi_{k+m} \\ 
 \[ \Psi_k , \Psi_m \] & \ =  \ & q (m-k) \, \Psi_{k+m} \\
 \[ \Phi_k , \Psi_m \] & \ =  \ & m p  \, \Psi_{k+m} \ - \ k q \, \Phi_{k+m}
\end{array}  \eqno(\sn.13)$$
and from these it follows that 
$$  \begin{array}{lll}
 \[ X_k , X_m \] & \ = \ & (m-k) \, X_{k+m} \\
 \[ X_k , Y_m \] & \ = \ &  m \, Y_{k+m} \\ 
 \[ Y_k , Y_m \] & \ = \ & 0 \ . \end{array} \eqno(\sn.14)$$
Notice that these are the same as those encountered in discussing the case {\bf S3}: we have thus to deal again with the Lie algebra $\G = \X \oplus_\to \Y$. Thus, provided we take into account remark \rnn, the algebraic computations considered there will immediately apply to this case as well. This correspondence between cases {\bf S3} and {\bf S4} is, of course, the one discussed in section 3 above.

\section{The S4 case: Poincar\'e renormalized forms}
\def\sn{9}

As remarked above, the algebra $\ker (\L_0 )$ is spanned by vector fields $\{ X_k , Y_k \}$ (with $k \in \N$) which generate the same Lie algebra $\G = \X \oplus_\to \Y$ encountered in discussing the case {\bf S3}, as also discussed in section 3. The fact that the
linear part is now given by $\zeta Y_0 = 2cpq Y_0$, rather than simply by $Y_0$, has no consequence on the discussion of linear subspaces, since the constants $c,p,q$ are all nonzero and thus $\zeta \not= 0$: see remark \rna.

We can thus just repeat the discussion conducted in case {\bf S3}, modulo remark {\rnn}  above; we write again $z := p + q$.

\subsection{General results}

We will thus consider the terms in $\W_z$, given by 
 $$ W_{z} = a_1
X_1 + b_1 Y_1 \ , \eqno(\sn.1)$$ 
 and consider the different cases
 $$
\cases{ 
 a_1 \not= 0 \ , \ b_1 = 0 \ ; & (a) \cr 
 a_1 = 0 \ , \ b_1 \not= 0
\ ;  & (b) \cr 
 a_1 \not= 0 \ , \ b_1 \not= 0 \ ; & (c) \cr 
 a_1 = 0 \ , \
b_1 = 0 \ . & (d) \cr } \eqno(\sn.2)$$
 
Notice that, considering a system which is already in standard normal form, the operators $\L_1 , ... , \L_{z-1}$ vanish; the first nontrivial higher homological operator is 
$$ \L_z := [ W_z , . ] \equiv [ a_1 X_1 + b_1 Y_1 , \, . \, ] \ .
\eqno(\sn.3)$$
 With exactly the same argument as in the discussion of the
case {\bf S3} we have the following results.
 
In case {\bf (a)}, where $W_z = a_1 X_1$, $\ker(\M_1) , ... , \ker
(\M_{z-1})$ do just coincide with $\ker (\L_0)$ (that is, the whole space on
which the trivial operators $\M_1 , ... \M_{z-1}$ are defined), while $\ker
(\M_z)$ reduces to the linear span of $W_0$ and $W_1$, i.e. of $Y_0$ and
$X_1$. The range of $\M_z$ is the whole
 linear span of the $\X , \Y$, except
the subspace spanned by $X_1 , X_2 , Y_1$. As vector fields $Z_1,Z_2$ which
are one in  $\ran (\M_z)$ and one in  $\ker (\M_1 )$ commute, no further
normalization is possible.

Thus, we obtain the same result as in case {\bf S3(a)}, with the role of
$\M_1$ now effectively played by $\M_z$ (which is the operator associated to
$X_1$, and more in general to $a_1 X_1 + b_1 Y_1$).

In cases {\bf (b)}, {\bf (c)} we do similarly reproduce the discussion of the
corresponding cases of {\bf S3}, again with the role of $\M_1$ now effectively
played by $\M_z$.

Thus we obtain the following expressions for the PRF when $W_z \not= 0$:
$$ \cases{ 
 \^W \ = \ \zeta Y_0 + a_1 X_1 + \^a_2 X_2 & (a) \cr 
 \^W \ = \ \zeta Y_0 + b_1 Y_1 + \sum_{k=2}^\infty \^a_k X_k & (b) \cr 
 \^W \ = \ \zeta Y_0 + (a_1 X_1 + b_1 Y_1 ) + \^a_2 X_2 & (c)
\cr}
\eqno(\sn.4)$$
In case {\bf (b)}, the $\G$-adapted procedure described in section 3  and subsection 6.3 will actually give a more reduced NF (the LRF), see below.

In the degenerate case {\bf (d)} we should again proceed as in case {\bf S3(d)}. With the same meaning for $\mu, \nu$ as there and the same splitting in
subcases {\bf (da)}, {\bf (db)}, {\bf (dc)}, we would obtain exactly the same expressions for the PRF as in (6.12), (6.16) and (6.20), at the exception of the linear part being given by $\zeta Y_0$ rather than by $Y_0$.

In case {\bf (db)}, using the $\G$-adapted procedure one would get, see (6.22), 
$$ \=W \ = \ \zeta Y_0 \, + \, a_\mu X_\mu \, + \, \^a_{2 \mu} X_{2 \mu} \, + \, \sum_{k=\nu}^\mu \, \^b_k Y_k \ ; \eqno(\sn.4') $$
this also applies to the case {\bf (b)}, with $\nu = 1$.

\subsection{Explicit computations}

The results of the explicit computations performed in the case {\bf S3} would
also extend to the present case. Indeed, once we have transformed the original
system into standard NF, the linear part $W_0$ does not enter in the PRF algorithm any more, and thus the presence of the constant $\zeta$ (instead than one) cannot affect the computations in any way. Again, when translating
the results obtained in case {\bf S3} to the present case, one has to take into account remark \rnn.

We will thus just follow the first steps of the computation in case {\bf (a)} to illustrate this.

We start from a system $W^{(1)}$ which has already been brought to 
standard
 NF. Transformations generated by $h_k \in \F_k$ for $0 < k < z$ are
necessarily trivial, as for such $k$ we have $\ker (\L_0) \cap \F_k = \{ 0
\}$, i.e. the Lie-Poincar\'e ``transformations'' reduce to the identity.  With a
transformation generated by $h_z = \a_1 X_1 + \b_1 Y_1$ the term $W_z$ is
unchanged, while $W_{2z}^{(z)}$ is taken into 
  $$ W_{2z}^{(z+1)} \ = \
a_2
 \, X_2 \ + \ (b_2 - a_1 \b_1 ) \, Y_2 \eqno(\sn.5)$$
  and we can of
course eliminate the
 $Y_2$ component by choosing $\b_1 = b_2 / a_1 $; we also
choose $\a_1 = 0 $.

Transformations with $h_k$, $z < k < 2z$ are trivial; thus we have
$W_m^{2z} \equiv W_m^{(z+1)}$ for all $m \ge 0$.
With a transformation generated by $h_{2z} = \a_2 X_2 + \b_2 Y_2 $ (so that
$h_{2z} \in \ker ( \L_0 ) \cap \F_{2z}$) the terms $W_k$, $k < 3z$, are
unchanged; the term $W_{3z}^{(2z)}$ is taken into 
 $$ W_{3z}^{(2z+1)} \ = \
[a_3 - a_1 \a_2 ] \, X_3 \ + \ 
 [ ( a_1 b_3 - a_2 b_2 - 2 a_1^2 \b_2 ) / a_1
] \, Y_3 \ . \eqno(\sn.6)$$
 This can be eliminated by choosing 
 $$ \a_2 \ = \ a_3 / a_1
\ \ , \ \ \b_2 \ = \ 
 (a_1 b_3 - a_2 b_2)/(2 a_1^2) \ . \eqno(\sn.7)$$

Again, the transformations with generator $h_k \in \ker(\L_0 ) \cap \F_k$ are
necessarily trivial for $2z < k < 3z$, and thus $W_m^{(3z)} = W_m^{(2z+1)}$
for all $m \ge 0$.

The explicit formulas obtained can be compared with those of section 6; they show the complete correspondence with the subcase {\bf S3(a)}. 
We believe there is no need to give further explicit formulas for the present case {\bf S4}, as they can be read from the corresponding ones for case {\bf S3}.
  
Notice that also the expression of the PRF in terms of the vector fields $X_k
, Y_k$ will be (except of course for the linear term $W_0$, where the constant
$\zeta$ appears) exactly the same as in the case {\bf S3}.
 
 \section{Summary of results for other cases}
\def\sn{10}

In this section we briefly recall, for the sake of completeness, the results 
obtained in \cite{LMP,IHP} for the other cases where the PRF is nontrivial.
These are cases {\bf S2} and {\bf N2} of our basic classification of linear parts.
An error contained in \cite{LMP,IHP} for one
degenerate {\bf S2} subcase is also corrected.

\subsection{PRFs and LRFs for the case S2}

In case {\bf S2} one could work in ${\bf C}^2$, but we will stay within the framework of the present discussion and work in $\R^2$, i.e. we will deal with real matrices, and thus write
$$ A \ = \ \pmatrix{ 0 & -1 \cr 1 & 0 \cr} \ . \eqno(\sn.1) $$
Notice that in order to map this case into {\bf S3}, it suffices to pass to polar coordinates.
It is well known that, with $r^2 = x^2 + y^2$, the standard NF is then 
$$ \begin{array}{rl}
{\dot x} \ =& \ - y \ + \ \sum_{k=1}^\infty r^{2k} ( a_k x - b_k y ) \\
{\dot y} \ =& \ \ x \ + \ \sum_{k=1}^\infty r^{2k} ( b_k x + a_k y )
\end{array} \eqno(\sn.2) $$

If the system is hamiltonian, then all the $b_k$ are zero; conversely, if all the $b_k$ are zero, the NF (2) is hamiltonian. Further reduction of the NF in
this case has been studied by Siegel and Moser \cite{SiM} a long time ago (their results have recently been shown to generalize to higher dimensions \cite{FoM}). We will consider the general (non-hamiltonian) case.

Let $\mu \ge 1$ be the smallest $k$ such that $a_k \not=0$, and $\nu \ge 1$ the
smallest $k$ such that $b_k \not=0$ (we assume both $\mu$ and $\nu$ are finite). 

If $\mu < \nu$, then (see subcases {\bf (a)} and {\bf (da)} for {\bf S3}) the PRF is given by 
$$ \begin{array}{rl}
{\dot x} \ = & \ - y \ + \ a_\mu r^{2\mu} x + \a r^{4\mu} x  \\
{\dot y} \ = & \ \ x \ + \ a_\mu r^{2\mu} y + \a r^{4\mu} y \ .  \end{array} \eqno(\sn.3)$$ 
If $\mu = \nu$, then (see subcases {\bf (c)} and {\bf (dc)} for {\bf S3}) the PRF is given by 
$$ \begin{array}{rl}
{\dot x} \ = & \ - y \ + \  r^{2\mu} (a_\mu x - b_\mu y) + \a r^{4\mu} x  \\
{\dot y} \ = & \ \ x \ + \  r^{2\mu} (b_\mu x + a_\mu y) + \a r^{4\mu} y \ .  \end{array} \eqno(\sn.4)$$
Here $a_\mu \not= 0$ and $b_\mu \not= 0$ are the same as in (2),
and the coefficient $\a$ is a real number. A detailed proof of this result is contained in section 12 of \cite{IHP}; a shorter proof is also given in \cite{LMP}.

If $\nu < \mu$, then (see cases {\bf (c)} and {\bf (dc)} for {\bf S3})
the PRF is given by 
$$ \begin{array}{rl}
{\dot x} \ = & \ - (1 + b_\nu r^{2 \nu} ) \, y \ + \ \sum_{k=\mu}^\infty r^{2k} \^a_k \, x  \\ 
{\dot y} \ = & \ \ (1 + b_\nu r^{2 \nu} ) x \ + \ \sum_{k=\mu}^\infty r^{2k} \^a_k \, y \ .  \end{array} \eqno(\sn.5)$$

In this same case, the LRF is given by  
$$ \begin{array}{rl}
{\dot x} \ = & \ - y \ + \ a_\mu r^{2\mu} x + \a r^{4\mu} x \ - \  \sum_{k=\nu}^\mu b_k \, y \\
{\dot y} \ = & \ \ x \ + \ a_\mu r^{2\mu} y + \a r^{4\mu} y \ + \ 
\sum_{k=\nu}^\mu b_k \, x .  \end{array} \eqno(\sn.6)$$ 
Here $a_\mu \not= 0$ and $b_k$ (for $\nu \le k \le \mu$) are the same as in (2), and the coefficient $\a$ is a real number. 

The computation for this case given in \cite{LMP,IHP} contained a mistake: the coefficients $b_k$ cannot be changed (via a PRF-like transformation) to eliminate the corresponding rotation terms without producing radial terms. It should also be stressed that the reduced NF obtained in these papers, even after correction of this mistake, is the LRF (and is not obtained with the generic PRF procedure); in particular, in the case $\nu < \mu$ this is {\it not } a PRF according to our definition.

It should be noticed that these results can be obtained, i.e. the case {\bf S2} can be studied, more easily using the approach of the present paper, as we now briefly indicate. 

We define, as in \cite{LMP,IHP}, dilation and rotation linear vector fields
$$ D = x \pa_x + y \pa_y \ \ , \ \ R = - y \pa_x + x \pa_y \eqno(\sn.7) $$
and with this compact notation, writing also $r^2 := (x^2 + y^2)$, we define
$$ \Psi_k \ := \ r^{2k} \, D \ \ , \ \ \Phi_k := r^{2k} \, R \ . \eqno(\sn.8)
$$ 

It is immediate to check that these vector fields satisfy the commutation relations
$$ [ \Psi_k , \Psi_m ] = 2 (m-k) \Psi_{k+m} \ , \ [ \Phi_k , \Phi_m ] = 0 \ , \ [ \Psi_k , \Phi_m ] = 2 m \Phi_{k+m} \ . \eqno(\sn.9) $$  
That is, we have the same algebraic structure as the one encountered in analyzing previous cases. We can make it identical, including coefficients, see (5.3),  by defining  $$ X_k = (1/2)^{1/3} \, \Psi_k \ \ , \ \ Y_k = (1/2)^{1/3} \, \Phi_k \ . \eqno(\sn.10) $$
In this way, the explicit computations performed for the case {\bf S3} can immediately be applied to this case as well. We write now the standard NF as $$ W \ = \ \zeta \, Y_0 \ + \ \sum_{k=0}^\infty (a_k X_k + b_k Y_k ) \eqno(\sn.11) $$
where $\zeta = (2)^{1/3}$. Formulas for the PRF can be read off the discussion of the {\bf S3} case. In particular, (6) corresponds to (6.22); see also (7.33).

\subsection{PRFs for the case N2}

In the (nonregular) case N2, we have
$$ A \ = \ \pmatrix{0&1\cr0&0\cr} \eqno(\sn.12) $$
and the standard NF is given by 
$$ \begin{array}{rl}
{\dot x} \ = & \ y + \sum_{k=1}^\infty b_k x^{k+1} \\
{\dot y} \ = & \sum_{k=1}^\infty (a_k x^{k+1} + b_k x^k y )
\end{array} \eqno(\sn.13) $$

Let $\mu \ge 1$ be the smallest $k$ for which $(a_k^2 + b_k^2) \not= 0$; then
the PRF is given by 
$$ \begin{array}{rl}
{\dot x} \ = & \ y  +  b_\mu x^{\mu+1} +  \a x^{\mu+2}
\ + \ \sum_{k=\mu+2}^\infty b_k x^{k+1} \\ 
{\dot y} \ = & \ a_\mu x^{\mu+1} + b_\mu x^\mu y + \a x^{\mu+1} y \ + \ 
\sum_{k=1}^\infty (a_k x^k y + b_k x^{k+1} ) \end{array} \eqno(\sn.14) $$
This represents a very poor simplification of the standard NF; such a poor performance of the algorithm is related to the vanishing of the semisimple part of $A$, i.e. to the fact the singular point is nonregular. 
A detailed proof of (14) is contained
in section 15 of \cite{IHP}. For a discussion of this singularity, see \cite{Tak} and \cite{BaS,KOW}.

\section{Conclusions}

We have studied vector fields in the plane around a singular point by means of normal forms theory, discussing in detail all possible cases when the singular point is a regular one. In doing this we have assumed the linearization of the vector field has been preliminarly taken into Jordan normal form. 

We have also shown that when the standard normal form (NF) is nontrivial, the Poincar\'e renormalized forms (PRFs) approach permits to substantially simplify the expression of the vector field in normal form. 

Thanks to constraints on the structure of the infinite dimensional Lie algebra of two-dimensional vector fields in normal forms \cite{Elp,Wal}, there is a substantial correspondence between different cases where the NF is nontrivial, and computations performed in one case can be mapped into any other one. We have taken advantage of this property, and performed explicit and detailed computations in one case ({\bf S3}), using them for other cases as well.

Considering the Lie algebra structure of vector fields in normal form also allow to define a different reduction scheme, designed to take advantage of this structure. The reduced normal form thus obtained, and called Lie renormalized form (LRF), is not necessarily a PRF; actually we have seen that in some of our subcases -- i.e. cases {\bf (b)} and {\bf (db)} for all the semisimple linear parts -- the LRF is finite while the PRF is infinite, and the LRF is not a special instance of PRF. The LRF approach is directly related to Broer's approach \cite{Bth,Bro} and to Baider's work \cite{Bai}.

The local behaviour of vector fields in $\R^2$ around regular singular points is of course very well studied, so that the real interest of our discussion is not in the expressions obtained themselves. 
Rather, we have shown that the PRF approach is viable to obtain a very explicit description of further reduced normal forms, even with the use of limited computing facilities.

The detailed analysis given here also led to implement considerations based on the Lie algebraic structure of vector fields in normal forms, and to define a $\G$-adapted reduction procedure, the LRF procedure, conjugating Lie-algebraic considerations {\it \`a la Broer } and the PRF algorithmic approach. This is of much wider use than the limited one considered here.

We have also corrected a computational error contained in previous work, and clarified (see also the appendix) some confusion on the issue of PRFs present in the literature.

\vfill\eject

\section*{Appendix. The Bruno system} 

As mentioned in remark 2 at the end of the Introduction, in his reviews of my papers \cite{LMP,IHP} for {\it Mathematical Reviews} \cite{Brep}, and again in his recent book \cite{Bru2} (section V.22) and in a preprint \cite{Bprep} which he was so kind to send me, A.D. Bruno has claimed that the main result of my works \cite{LMP,IHP} is  wrong; he also gave a ``counterexample'' to my result. 

This claim seems to originate in a misunderstanding of my previous papers, and the purpose of this appendix is to clarify the situation, also by completely explicit computations, dealing precisely with the system proposed by Bruno.

The ``counterexample'' mentioned above falls in subcase {\bf S3(b)} of the classification considered here and was given  in \cite{Brep,Bru2} as
$$ \begin{array}{rl}
 {\dot x} \ = \ & x^3 \\
{\dot y} \ = \ & y + xy + x^2 y \ \equiv \ (1+x+x^2) y \ ; \end{array} 
\eqno(A.1) $$
according to Bruno \cite{Brep}, the PRF for this
system would be given by
$$ \begin{array}{rl}  {\dot x} \ = \ & x^3  \\  
{\dot y} \ = \ & y + xy  \ \equiv \ (1+ x) y \end{array}  
 \eqno(A.2') $$
with no higher order terms. In \cite{Bru2}, this is changed to 
$$ \begin{array}{rl}  {\dot x} \ = \ & a_2 x^3 + \a x^5 \\  
{\dot y} \ = \ & y + \b xy  \ \equiv \ (1+ \b x) y \ , \end{array}  
 \eqno(A.2'') $$
again with no higher order terms, where $a_2 , \a , \b$ are some constants (no mention is given to this discrepancy in \cite{Bru2}; in both cases no computation is reported to explain how these are obtained). More recently, Bruno and Petrovich  \cite{Bprep} also considered a slightly generalized form of this ``counterexample'', see below.
 
The discussion of section 5, and the explicit computations of section 6, show that (A.2) -- in either one of its versions -- is {\it not } the PRF for system (A.1). 

Actually, in his reviews, book, and preprint, Bruno quotes my result in a form which does not correspond to -- and is not equivalent to -- the statements I gave in \cite{LMP,IHP}; thus his sweeping assertion that ``it is easy to see that the statement of Gaeta is wrong'' (see p. 275, \cite{Bru2}) does refer to an incorrectly reported version\footnote{Notice that according to Bruno's definition (but with the notations of the present paper) the PRF would be characterized by the property that $\L_{j}^+ (W_k ) = 0$ for $j < k$ in the first review \cite{Brep}, and for $j = k-1$ in the second of \cite{Brep} and in \cite{Bru2}.} of my result. 

The key difference is given by the fact that the role played by the $\M_k$ operators in my construction is, in the version reported by Bruno \cite{Brep,Bru2,Bprep}, taken by the $\L_k$ ones (with no restriction to kernels of $\L_s$ with $s < k$; see section 2). Thus, the result ``quoted'' by Bruno should not be attributed to my papers  (incidentally, the statement attributed to me in \cite{Brep,Bru2,Bprep} is wrong in an obvious way). 
 
To avoid any confusion about PRF for the system (A.1), let us  
perform the Poincar\'e renormalization algorithm up to terms in $\W_9$. I stick to the proper general PRF procedure as stated in \cite{LMP,IHP}, i.e. not consider the LRF scheme (actually the system (A.1) is already in LRF). I will freely use the notation introduced in discussing the case {\bf S3}.

\bigskip

We write $W = x^3 \pa_x + y(1+x+x^2) \pa_y $.  The system is already in NF, so we write $W^{(1)} = W$; the system is taken into PRF by operating successive
transformations with generators $H_k = \a_k X_k + \b_k Y_k$ (to avoid any possible misunderstandings, let us specify there is no sum on $k$). 
 
It results that choosing $\b_k = 0$ and with  
$$ \begin{array}{l}
\a_1  = \ -1 \ , \ 
 \a_2  = \ 1  \ , \ 
 \a_3  = \ -2 \ , \ 
\a_4  = \ 9/2 \ , \\ 
\a_5  = \ - 12 \ , \ 
\a_6  = \ 33  \ , \ 
\a_7  = \ - 99 \end{array} \eqno(A.3)$$
the system takes successively the forms (where $O(9)$ denotes terms in  $\W_9$ and higher)
$$ \begin{array}{rl}
\=W^{(2)} \ = & \ Y_0 + Y_1 + X_2 - X_3 - Y_3 + X_4 + 2 Y_4 + \\
 & - X_5 - 3 Y_5 + X_6 + 4 Y_6 - X_7 - 5 Y_7 + X_8 + 6 Y_8 + O(9) \ ; \\
\=W^{(3)} \ = & \ Y_0 + Y_1 + X_2 - X_3  +
X_4 + 2 Y_4 - 2 X_5 - (9/2) Y_5 + \\ 
 & + 3 X_6 + 12 Y_6 - (11/2) X_7 - 
25 Y_7 + 9 X_8 + 54 Y_8 + O(9) \ ; \\ 
\=W^{(4)} \ = & \ Y_0 + Y_1 + X_2 - X_3  + X_4 - (9/2) Y_5 + \\ 
 & + 3 X_6 + 12 Y_6 - (15/2) X_7 - 
33 Y_7 + 13 X_8 + 99 Y_8 + O(9) \ ; \\
\=W^{(5)} \ = & \ Y_0 + Y_1 + X_2 - X_3  + X_4 - 6 X_6 + \\ 
 & + 12 Y_6 - 3 X_7 - 33 Y_7 
+ 13 X_8 + 99 Y_8 + O(9) \ ; \\ 
\=W^{(6)} \ = & \ Y_0 + Y_1 + X_2
- X_3  + X_4 
 - 6 X_6  + 33 X_7 + \\ 
 & - 33 Y_7 
- 11 X_8 + 99 Y_8 + O(9) \ ; \\
\=W^{(7)} \ = & \ Y_0 + Y_1 + X_2 - X_3  + X_4 
- 6 X_6  + 33 X_7  
- 143 X_8 + 99 Y_8 + O(9) \ ; \\
\=W^{(8)} \ = & \ Y_0 + Y_1 + X_2 - X_3  + X_4 
- 6 X_6  + 33 X_7 - 143 X_8  + O(9) \ . \end{array} \eqno(A.4)$$
The latter is the PRF, up to terms $O(9)$, for (A.1).

In \cite{Bprep}, Bruno and Petrovich consider general systems with linear part corresponding to our case {\bf S3}. The standard normal form for these is, as discussed above, of the form
$$ \begin{array}{rl}
{\dot x} \ = & \ \sum_{k=1}^\infty \, a_k \, x^{k+1} \\
{\dot y} \ = & \ y \, + \, \sum_{k=1}^\infty \, b_k \, x^k y \ ; \end{array}
\eqno(A.5)$$
we denote by $m \ge 1$ the smallest $k$ such that $a_k \not=0$, and by $\ell \ge 1$ the smallest $b_k$ such that $b_k \not=0$. That is, we assume $a_k = 0$ for $k < m$, and $b_k = 0 $ for $k < \ell$. 

Bruno and Petrovich suggest to consider the case $\ell \le m < \infty$, and give as an example the case $\ell = 1$, $m=2$.
They claim, see formula (5.2) of \cite{Bprep}, that the PRF in this case is
$$ \begin{array}{rl}
{\dot x} \ = & \ a_m x^{m+1} \, + \, \a x^{2m+1} \\
{\dot y} \ = & \ y \, + \, \b x^\ell y \ ; \end{array}
\eqno(A.6)$$
with $a_m \not= 0$, and $\a,\b$ real coefficients.

Once again this formula does not correspond to the PRF computed in above, so that their proof that the original system (A.5) cannot be conjugated -- even formally -- to system (A.6), does not
concern PRFs. 

\bigskip

Let us now consider the domain of convergence of the normalizing transformation explicitly considered above. Here $\b_k = 0$ for all $\b$, so that the mappings do not act on $y$. It is easy to obtain (with the help of an algebraic manipulation program) explicit expressions for the changes of coordinates and thus for the domain of analyticity of the overall transformation up to step $k$.

We write the combined effect of the first $k$ changes of coordinates as $x \to \=x^{(k)} = x / B_k (x)$; the denominators $B_k (x)$ can
be written in recursive terms as
$$ B_k (x) \ = \ \[ \( B_{k-1} (x) \)^k \, - \, (-1)^k \, \gamma_k \, x^k \]^{1/k} \ . \eqno(A.7)$$
The first numbers of the sequence $\gamma_k$ ($k=1,2,...$) are given by 1, 2, 6, 18, 60, 198, 693; obviously $B_0 (x) \equiv 1$. We will omit the derivation of this recursion formula.

This also allow to explicitely determine the domain of analiticity of the transformations, which can be read from the roots of $B_k (x)$; but the expression so obtained quickly become  extremely involved and of little interest. 

Thus I have computed analytically the $x_-^{(k)} , x_+^{(k)} $ such that the overall transformation up to step $k$ (i.e. $x \to x^{(k)}$) is analytic in the strip $x_-^{(k)} < x < x_+^{(k)}$, but just report here their numerical value. These are:
$$ \begin{array}{l}
x_-^{(1)} = -1 \ , \ x_-^{(2)} = -0.333333 \ , \ x_-^{(3)} = -0.270929 \ , \ x_-^{(4)} = -0.244594 \ , \\ 
x_-^{(5)} = -0.228796 \ ,  x_-^{(6)} = -0.21915 \ , \ x_-^{(7)} = -0.21224 \ ; \\
x_+^{(1)} = \infty \ , \ x_+^{(2)} = 1. \ , \ x_+^{(3)} = 1. \ , \ x_+^{(4)} = 0.668534 \ , \\  
x_+^{(5)} = 0.668534 \ , \ x_+^{(6)} 0.561419 \ , \ x_+^{(7)} =  0.561419 \ . 
\end{array} \eqno(A.8)$$

\bigskip

Finally, let us clarify the relation of the computations presented in this appendix with the problems, mentioned above, concerning the discussion of case ${\bf S2(db)}$) with $\nu < \mu$ in \cite{LMP,IHP}.
 
It should be remarked that if one maps (as discussed in section 3, or simply passing to polar coordinates) case {\bf S2} to case {\bf S3}, then the wrong result given in \cite{LMP,IHP} for the degenerate case (with the notation of the present paper, ${\bf S2(db)}$) with $\nu < \mu$ would map exactly to the ``pseudo-PRF'' (A.2') given by Bruno in \cite{Brep}. Notice however that there it is claimed that the error lies with the general statement and not with the computations of the example; also, (A.2') is not claimed to be derived from the results for {\bf S2} given in \cite{LMP,IHP}, but just to be the PRF according to the definition reported there. Thus, such a statement appears somehow mysterious; however, as discussed above, it does not involve PRFs according to their definition given in \cite{LMP,IHP} and in this paper, so that we don't have to deal with it.

\vfill\eject

\end{document}